\documentclass{emulateapj}
\slugcomment{draft v1}

\shorttitle{Electric field screening at polar cap}
\shortauthors{Kisaka, Asano \& Terasawa}

\begin{document}


\title{Electric field screening with back-flow at pulsar polar cap}


\author{Shota Kisaka\altaffilmark{1,2}}
\email{kisaka@phys.aoyama.ac.jp}
\author{Katsuaki Asano\altaffilmark{3}}
\email{asanok@icrr.u-tokyo.ac.jp}
\author{Toshio Terasawa\altaffilmark{3,4}}
\email{toshio.terasawa@riken.jp}


\altaffiltext{1}{Department of Physics and Mathematics, Aoyama Gakuin University, Sagamihara, Kanagawa, 252-5258, Japan}
\altaffiltext{2}{JSPS Research Fellow}
\altaffiltext{3}{Institute for Cosmic Ray Research, The University of Tokyo, Kashiwa, Chiba, 277-8582, Japan}
\altaffiltext{4}{iTHES Research Group, RIKEN, Wako, Saitama 351-0198, Japan}


\begin{abstract}

Recent $\gamma$-ray observations suggest that the particle acceleration occurs at the outer region of the pulsar magnetosphere. The magnetic field lines in the outer acceleration region (OAR) are connected to the neutron star surface (NSS). If copious electron--positron pairs are produced near the NSS, such pairs flow into the OAR and screen the electric field there. To activate the OAR, the electromagnetic cascade due to the electric field near the NSS should be suppressed. However, since a return current is expected along the field lines through the OAR, the outflow extracted from the NSS alone cannot screen the electric field just above the NSS. In this paper, we analytically and numerically study the electric-field screening at the NSS taking into account the effects of the back-flowing particles from the OAR. In certain limited cases, the electric field is screened without significant pair cascade if only ultrarelativistic particles ($\gamma\gg1$) flow back to the NSS. On the other hand, if electron--positron pairs with a significant number density and mildly relativistic temperature, expected to distribute in a wide region of the magnetosphere, flow back to the NSS, these particles adjust the current and charge densities, so that the electric field can be screened without pair cascade. We obtain the condition for the number density of particles to screen the electric field at the NSS. We also find that in ion-extracted case from the NSS, bunches of particles are ejected to the outer region quasi-periodically, which is a possible mechanism of observed radio emission.
  
\end{abstract}





\section{INTRODUCTION}

\label{intro}

In pulsar magnetosphere, particles are significantly accelerated at the given regions, 
and emit electromagnetic radiation from radio to $\gamma$-ray wavelength. 
Observations by {\it Fermi Gamma-Ray Space Telescope} have shown that the differential spectra above 200 MeV 
are well described by the power-law functions with an exponential cut-off, 
and that the cutoff shapes sharper than the simple exponential cutoff are rejected with high significance 
\citep[e.g., ][]{vela09}. 
This rules out the near-surface emission proposed in the polar cap cascade model \citep{DH96}, 
which would exhibit a much sharper spectral cut-off due to the attenuation of the magnetic pair-creation. 
Hence, the detected $\gamma$-ray pulse emission should originate from the outer region of the magnetosphere, 
as considered in the outer gap model \citep[e.g., ][]{CHR86, R96, THSC06, H06, H15, TNC15},
as well as the current sheet model \citep[e.g., ][]{KSG02, BS10, KHK14, BKHK15, CPS15}.

On the other hand, the region just above the neutron star surface (NSS) has been considered 
as the site of the radio pulsed emission \citep[e.g., ][]{Nou+15}. 
The mechanism of pulsar radio emission is established as a coherent process, so that the plasma dynamics near the NSS
would be strongly related to the emission mechanism \citep[e.g., ][]{S71}.
The two-stream instability is a promising process to create the plasma bunches
\citep[e.g., ][]{RS75}. 
The curvature radiation from the bunches is usually discussed as the mechanism of the coherent radio emission \citep[e.g., ][]{S75}.
In order to investigate the possible instabilities near the NSS, 
we should take into account the non-stationary effects in the plasma flows.

The dynamics of the plasma and the electromagnetic field near the NSS highly depends on the ratio of the current density 
parameter along the magnetic field, $j_{\rm m}$, to the Goldreich-Julian (GJ) value, 
$j_{\rm GJ}=\rho_{\rm GJ}c$ \citep{Me85, S97, TA13}, which is characterized by the GJ charge density 
$\rho_{\rm GJ}=-{\bf \Omega}\cdot{\bf B}/2\pi c$ \citep{GJ69}, 
where ${\bf \Omega}$ is the stellar angular velocity vector and ${\bf B}$ 
is the local magnetic field vector. 
The parameter $j_{\rm m}$ is regulated by the twist of the magnetic field 
($\nabla\times{\bf B}$) imposed by the global stress balance of the pulsar magnetosphere \citep[e.g., ][]{S91}.
In the polar cap region near the NSS, an accelerating electric field spontaneously develops to adjust
the current and charge densities to the current density parameter $j_{\rm m}$ and the GJ charge density $\rho_{\rm GJ}$.
In the cases $j_{\rm m}/j_{\rm GJ}\le0$ and $j_{\rm m}/j_{\rm GJ}\ge1$, outflowing particles from the NSS alone cannot 
adjust the current and charge densities to $j_{\rm m}$ and $\rho_{\rm GJ}$ simultaneously \citep[e.g., ][]{Me85}.
In such cases, a significant accelerating electric field develops and causes the copious pair creation.
The newly created pairs would screen the accelerating electric field
for a temporary period of time \citep[e.g., ][]{S71, LMJL05, TA13}.

The back-flowing particles from the outer acceleration region (OAR)
modify the above description of the dynamics near the NSS. 
As a result of discharge at the OAR in the magnetosphere
(e.g., outer gap or current sheet), 
some fraction of charged particles would come back to the NSS. 
Such back-flowing particles are actually seen in numerical studies \citep[e.g., ][]{H06, WS07, CB14, CPPS15}.
The existence of the back-flow is favorable to explain the observed pulse profiles
in the non-thermal soft $\gamma$-ray, X-ray and optical wavelengths, 
whose peaks are not aligned with the GeV $\gamma$-ray one \citep{TCS08, KK11, WTC13}. 
The back-flowing particles have been also considered to heat the NSS around the magnetic pole, 
whose signature is observed as the thermal pulsed emission in soft X-ray band \citep[e.g., ][]{ZP04}.
The threshold of the occurrence of pair cascade near the NSS depends on the contribution of the 
back-flowing particles to the current and the charge densities \citep{B08}.
The outflow from the NSS would also affect the dynamics in the OAR.
The outgoing particles from the NSS contribute to the particle injection rate from the inner boundary of the OAR
\citep[e.g., ][]{THSC06}.

If an outflow from the NSS affects the accelerating electric field in the OAR, 
the current and number densities of the back-flowing particles change, 
and the resulting particle outflow from the NSS may be also modified. 
Through such non-linear interplay between the NSS and OAR, 
the global magnetosphere is expected to reach the steady or quasi-steady state (e.g., periodic behavior). 
\citet{Le+14} and \citet{TNC15} suggest that a non-stationary outer gap model is favored 
to reproduce sub-exponential cut-off feature in the GeV $\gamma$-ray spectrum observed with {\it Fermi}. 
In order to understand the global behavior of the magnetosphere, 
we need to link the dynamics of the region above the NSS and the OAR.

In the first step,
we focus on the dynamics of only the restricted region just above the NSS for given back-flowing particles. 
If a steady electric field just above the NSS exists,
copious electron--positron pairs are produced via electromagnetic cascade. 
Such pairs flow into the OAR, and may screen out the electric field in the OAR.
Therefore, the electric field just above the NSS should be almost screened out to activate the OAR.

The local simulations have been performed to investigate the particle acceleration 
and the pair creation processes near the NSS \citep{BT07, L09, T10, B11, CB13, TA13, TH15}.
Since the present global simulations are difficult to include the realistic pair-creation process 
with the actual mass-to-charge ratio \citep{SA02, WS07, WS11, YS12, PS14, CB14, CPPS15, B15, PSC15, PCTS15}, 
the local simulations are complementary.
In order to link the local region above the NSS to the global magnetospheric structures,
it is useful to model the properties of the outflowing particles from the NSS for arbitrary ratio
$j_{\rm m}/j_{\rm GJ}$ and the back-flow from the OAR.

\citet{T10}, \citet{TA13} and \citet{TH15} performed
the local particle simulations to investigate the pair cascade near the NSS.
In their regimes, a large number density of pairs are supplied to the OAR,
so that the electric field in the OAR is screened by the copious pairs.
Then, the back-flow from the OAR would be suppressed. 
In this context, 
the effect of the back-flowing particle has not been investigated in the local particle simulations so far.
However, the OAR as a source of the back-flowing particles should
exist if the pair cascade near the NSS fails to supply enough pairs.

In this paper, we study the screening of the accelerating electric field above the NSS
taking into account the effect of the back-flowing particles from the OAR. 
As we have mentioned above, we consider that the screening of the electric field near the NSS is
a necessary condition to activate the OAR, because too much pair supply from the inner magnetosphere
via a strong electric field would choke the OAR.
The local condition of electric field screening near the NSS in the absence of the pair cascade is investigated
for a given ratio, $j_{\rm m}/j_{\rm GJ}$, which is imposed in the global magnetospheric structure. 
In Section \ref{model}, we introduce our model with a particle outflow from the NSS, 
where the number density and momentum distribution of the back-flowing particles are given as model parameters. 
In Section \ref{analytic}, we analytically show the screening condition for the velocity of the plasma flow from the NSS 
in the case where 
the back-flowing particles are ultra-relativistic ($\gamma\gg1$). 
We see that the development of the accelerating electric field cannot be avoided 
for some combinations of the total current density and the contribution of the back-flowing particles. 
In Section \ref{numerical}, we introduce additional components of back-flowing particles, 
electron--positron pairs with a mildly relativistic temperature.  
Particle-in-Cell simulations are performed to investigate the screening conditions near the NSS. 
Implications of our results for the pulsar radio emission is discussed in Section \ref{radio}.
We summarize our work in Section \ref{summary}.

\section{MODEL} 
\label{model}

We consider a local problem of how the accelerating electric field is screened near the NSS 
for given parameters. 
This study is motivated by the idea that the field screening near the NSS
may be an essential condition to activate the OAR,
from which high-energy photons are emitted.
Our model is one dimensional (1D), with spatial axis along magnetic field lines. 
We assume that charged particles move along straight magnetic field lines, which are perpendicular to the NSS. 
This assumption is justified by following reasons. 
One is that charged particles in the strong magnetic field
are in the first Landau level and move strictly along magnetic field lines. 
The other is that the length scale $L$ of our calculation domain 
is much smaller than the radius of the field line curvature $R_{\rm cur}$ and the radius of the polar cap $r_{\rm pc}$ 
\footnote{For $\gamma\gg10^6$, we cannot neglect the effect of curvature radiation.
  However, we do not treat such a high-energy particle in most cases.}. 
We neglect the induced variations of the magnetic field
that accompany variable electric field parallel to the magnetic field, $E_{\parallel}$. 

For the boundary condition on the NSS, 
we assume that electrons or ions with non-relativistic velocity are freely extracted from the NSS 
by the electric field $E_{\parallel}$ just above the NSS. 
This assumption is reasonable because the work function is small as compared with thermal energy for most pulsars
\citep{J80, M84, NLK86}. 

One of the most important parameters is the magnetospheric current density parameter
$j_{\rm m}=(c/4\pi)\nabla\times{\bf B}$.
The parameter $j_{\rm m}$ is induced to balance the global stress, since open magnetic field lines 
that pass through the light cylinder are twisted.
When the actual current density $j$ coincides with $j_{\rm m}$, the electric field becomes stationary.
However, the condition $j=j_{\rm m}$ is not always assured. 
In general, $j_{\rm m}$ can take any values, depending on some global conditions,  
because the local accelerator and the pulsar wind interplay each other through the current. 
Thus, we cannot deduce the value of $j_{\rm m}$ from any local model.
In our local model, we regard the value of $j_{\rm m}$ as a model parameter.

The local current density, $j$, tends to be quickly adjusted to $j_{\rm m}$ via the generation of the accelerating 
electric field $E_{\parallel}$. 
As a result, the character of the accelerator near the NSS strongly depends on $j_{\rm m}$
\citep{Me85, S97, B08, TA13}.
Since the parameter $j_{\rm m}$ should have the magnetospheric timescale, it can be regarded as stationary 
on the dynamical timescale typical for the surface region. 
which is much longer than the local time scales $L/c$ or $r_{\rm pc}/c$. 
We assume that the imposed parameter $j_{\rm m}$ is constant,
while the local electric field and plasma flows develop with time \citep{T10, TA13}.

\begin{figure}
 \begin{center}
  \includegraphics[width=80mm]{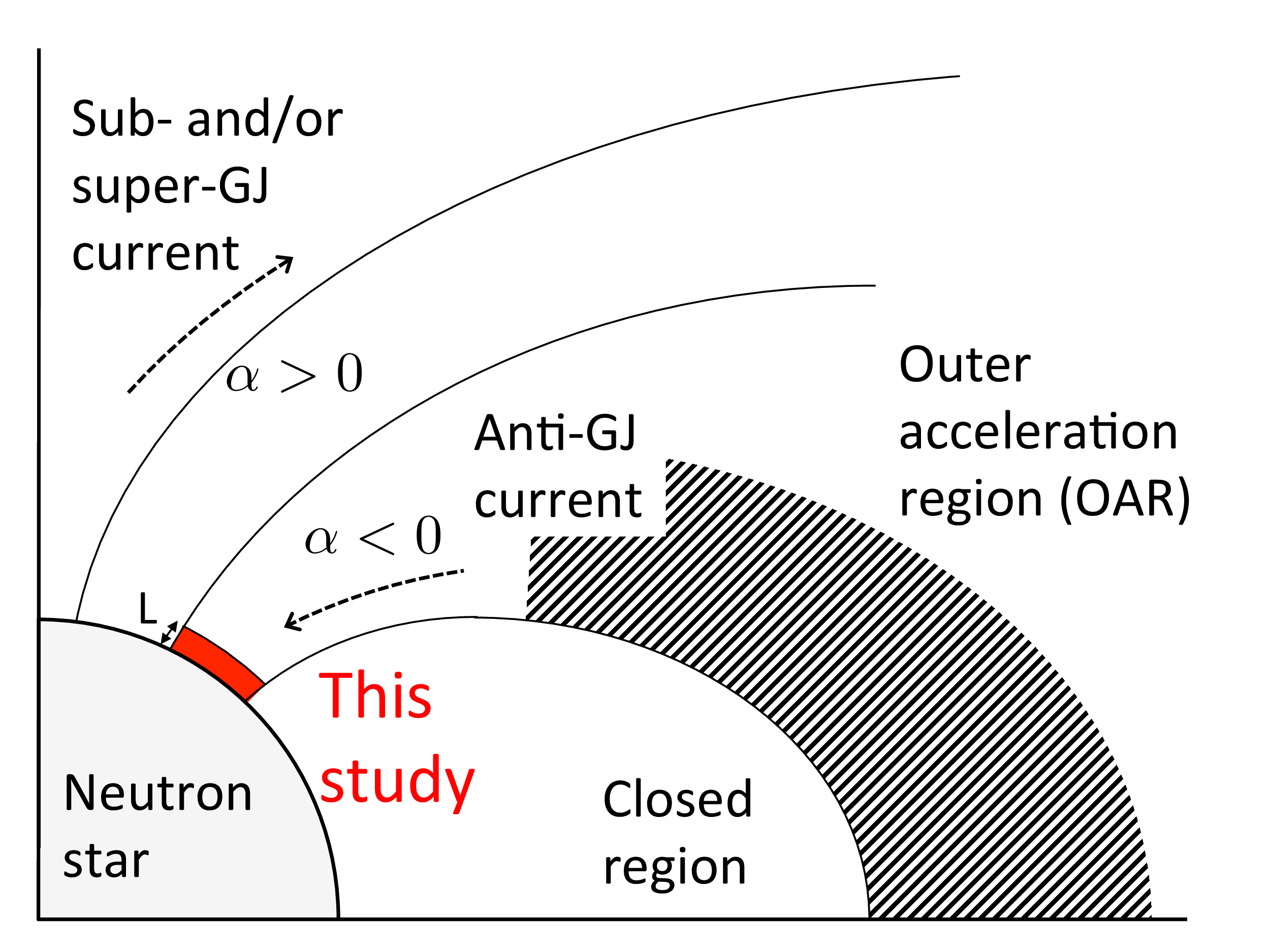}
 \caption{Schematic picture of the current closure in the pulsar magnetosphere. 
We define the dimensionless current density $\alpha\equiv j/\rho_{\rm GJ}c$, 
where $\rho_{\rm GJ}$ is the GJ charge density at the polar cap surface. 
The shaded area denotes the OAR. 
The length scale $L$ is our calculation domain. 
The current flowing in the OAR implies the anti-GJ ($\alpha<0$). }
 \label{schematics}
 \end{center}
\end{figure}

The main targets in this paper are the effects of the back-flowing particles from the OAR
on the plasma dynamics near the NSS (Figure \ref{schematics}). 
Thus we consider the energetic pulsars whose OAR is sustained by the pair cascade.
If electron--positron pairs whose number density is much larger than the GJ value, $n_{\rm GJ}=|\rho_{_{\rm GJ}}|/e$, 
are flowing back to the NSS, the electric field would be easily screened \citep[e.g., ][]{L09}. 
Here, we focus on the cases where the number density of the back-flowing particle is an order of the GJ value. 
For a pulsar with active OAR, this is naturally expected as follows.
The OAR locates on the last-open field line
which is the boundary between the open and the closed field lines
\citep[shaded region in Figure \ref{schematics}; e.g., ][]{CHR86}.
The returning current ($j/\rho_{\rm GJ}c<0$) should flow along the field lines in the OAR to close the current circuit 
(Figure \ref{schematics}) \footnote{Note that for a nearly orthogonal rotator or an inclined  millisecond pulsar, their polar cap may cross the equatorial plane. Then, the current $j/\rho_{\rm GJ}c$ along the field lines in the OAR would be positive at the surface. However, the polar cap radius is much smaller than the stellar radius so that we can expect a negative value for $j/\rho_{\rm GJ}c$ for most normal pulsars.}. 
The particles accelerated in the OAR would dominantly contribute to the return current,
so that the expected value of the returning current density is an order of the GJ value. 
The typical current density in the results of the global simulaitons is an order of GJ value \citep[e.g., ][]{Kal12}.

In addition to the accelerated particles, the electron-positron pairs created between the OAR 
and NSS are also expected to come back to the NSS.
Recent time-dependent particle simulations show the existence of two components 
in the momentum distribution, which stationary screen the electric field $E_{\parallel}$ outside the OAR 
\citep{T10, TA13}. 
A beam component with high Lorentz factor accounts for the current density as $j=j_{\rm m}$, while 
another component with a quasi-thermal momentum distribution and an average velocity 
$\bar{\beta}\sim 0$ adjusts the charge density as $\rho=\rho_{\rm GJ}$ \citep{T10, CB13, TA13}. 
In the outer gap model, the electric field $E_{\parallel}$ outside the OAR is screened 
via photon-photon pair creation near the null-charge surface \citep{CHR86, THSC06, H06}. 
Most pairs are created above the OAR, because the curvature photons are emitted to
the tangential direction of the field line.
On the other hand, at the inner boundary of the OAR, 
the number density of created pairs may not be much larger than the GJ value.
When the screening process with the two plasma components works near the null-charge surface, 
the non-relativistic particles are also expected to flow towards the NSS from the OAR.
In what follows, we consider two cases for the momentum distribution of the back-flowing particles
with a comparable density to the GJ number density. 

\begin{figure}
 \begin{center}
  \includegraphics[width=80mm]{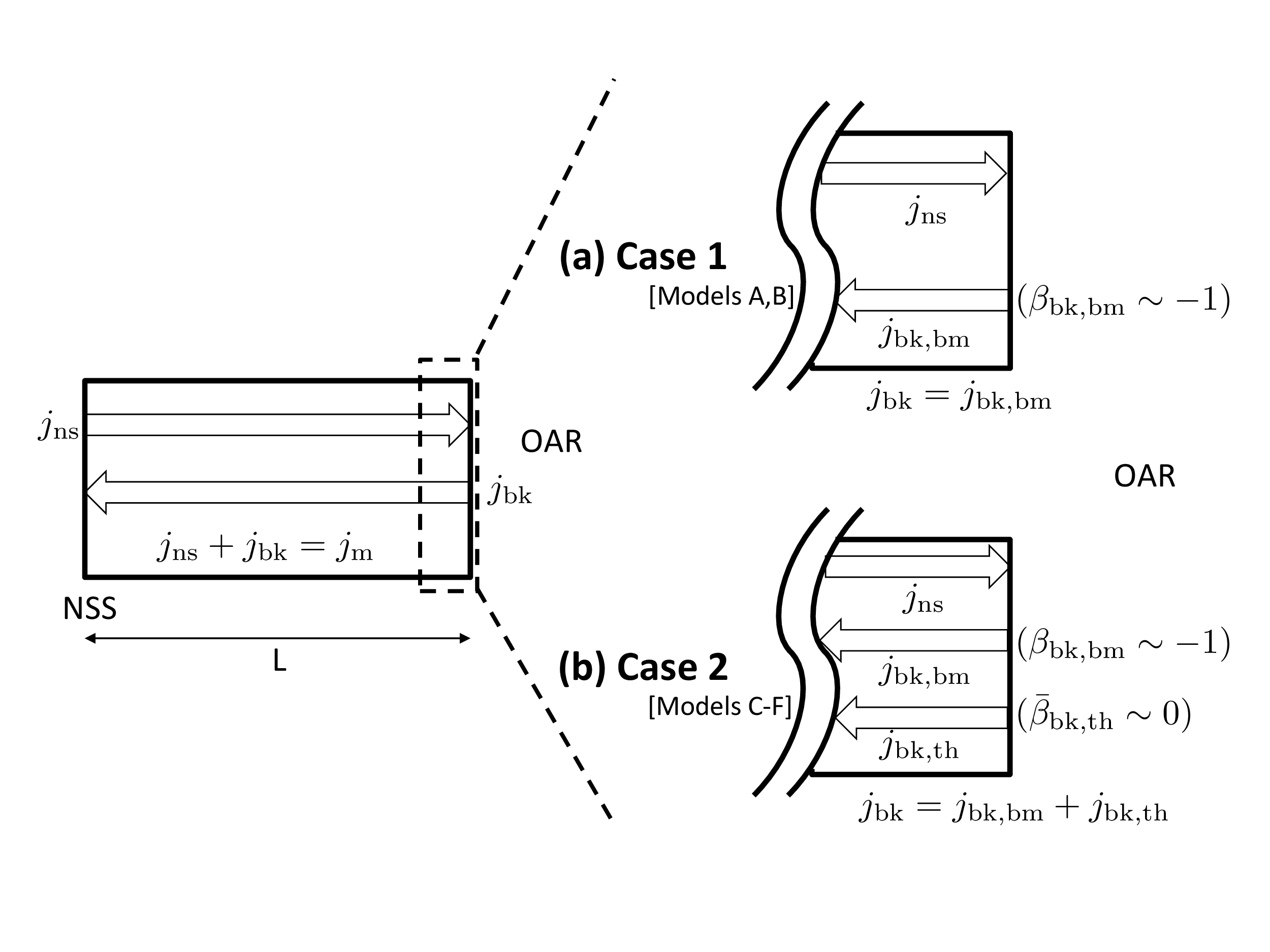}
 \caption{The schematic pictures of the current flows in the calculation domain with the screening condition
   of the electric field for the currents. 
Panels (a) and (b) are enlarged view near the outer boundary. 
Panel (a); only particles with $\gamma\gg1$ are flowing back to the NSS. 
Panel (b); both particles with $\gamma\gg1$ and $\gamma\sim1$ are flowing back to the NSS.}
 \label{image}
 \end{center}
\end{figure}

In case 1, we assume that only particles with large Lorentz factor $\gamma \gg 1$ are flowing back to the NSS. 
The schematic pictures of the case with a back-flow are shown in Figure \ref{image} (a).
The velocity of the back-flowing particles is almost light speed until reaching the NSS. 
Hereafter, we call these particles a beam component. 
The beam component consists of electrons or positrons, and is accelerated in the OAR.
The number density of the beam component would be an order of GJ value, $n_{\rm GJ}$.
The Lorentz factor $\gamma$ is regulated by the acceleration process in the OAR and the energy loss 
due to the curvature and synchrotron radiations during the travel to the NSS. 
As seen in the numerical results by \citet{H06}, particles
may be slightly decelerated before reaching the NSS ($\gamma\sim10^5-10^6$). 
We neglect the particle creation by the curvature photons emitted by the beam component. 
We analytically investigate the screening condition for case 1 in Section \ref{analytic}.

In case 2, we consider the possibility that there is another component of
the back-flowing particles with a quasi-thermal momentum distribution and an average 
velocity $\bar{\beta}\sim0$ [see Figure \ref{image} (b)]. 
Hereafter we call this second component a thermal component. 
The number density of the thermal component is assumed to be an order of $n_{\rm GJ}$ as we mentioned. 
For case 2, we study the screening of $E_{\parallel}$
with kinetic time-dependent particle simulations in Section \ref{numerical}.

\section{ANALYTICAL DESCRIPTION FOR SCREENING CONDITIONS OF ELECTRIC FIELD}
\label{analytic}

Here, we analytically describe the screening conditions in 1D model. 
The momentum dispersion of the flows is neglected in this section.
Note that the results do not depend on the sign of the GJ charge density $\rho_{\rm GJ}$.
Hereafter, we denote a quantity $Q$ due to the outflow from the NSS as $Q_{\rm ns}$, 
and that due to the back-flow as $Q_{\rm bk}$.
 
Let us introduce dimensionless current densities for each component $j_{\rm k}$ 
(e.g., $j_{\rm ns}$ and $j_{\rm bk}$ in Figure \ref{image}) and the current density parameter $j_{\rm m}$, as
\begin{eqnarray}\label{alpha}
\alpha_{\rm k}=\frac{j_{\rm k}}{c\rho_{\rm GJ}},~~~~~
\alpha_{\rm m}=\frac{j_{\rm m}}{c\rho_{\rm GJ}}.
\end{eqnarray}
We define the charge density for each component, $\rho_{\rm k}$. 
The dimensionless average velocity of a current component, 
$\beta_{\rm k} = j_{\rm k}/\rho_{\rm k}c$, relates to $\alpha_{\rm k}$ as 
\begin{equation}\label{betaByAlphaRhoRhoGJ}
\displaystyle{
 \beta_{\rm k}= \frac{\alpha_{\rm k}}{ (\rho_{\rm k}/\rho_{\rm GJ}) }.
}
\end{equation}
The positive (negative) direction of $\beta_{\rm k}$ corresponds to the outward direction to (inward direction from) the magnetosphere. 

In the 1D case, the electric field satisfies Maxwell's equations that lead to \citep[e.g., ][]{LMJL05} 
\begin{eqnarray}\label{poisson}
\nabla\cdot  E_{\parallel} = 4\pi(\rho-\rho_{\rm GJ}),
\end {eqnarray}
\begin{eqnarray}\label{ampere}
\frac{\partial E_{\parallel}}{\partial t}=4\pi c\rho_{\rm GJ}( \alpha_{\rm m}- \alpha),
\end {eqnarray}
in the co-rotating frame of the star, where $\rho\equiv\Sigma_{\rm k}\rho_{\rm k}$ and 
$\alpha\equiv\Sigma_{\rm k}\alpha_{\rm k}$ are the total charge and dimensionless current densities, respectively. 

\begin{figure*}
 \begin{center}
 \hspace{-30mm}
  \includegraphics[width=90mm, angle=270]{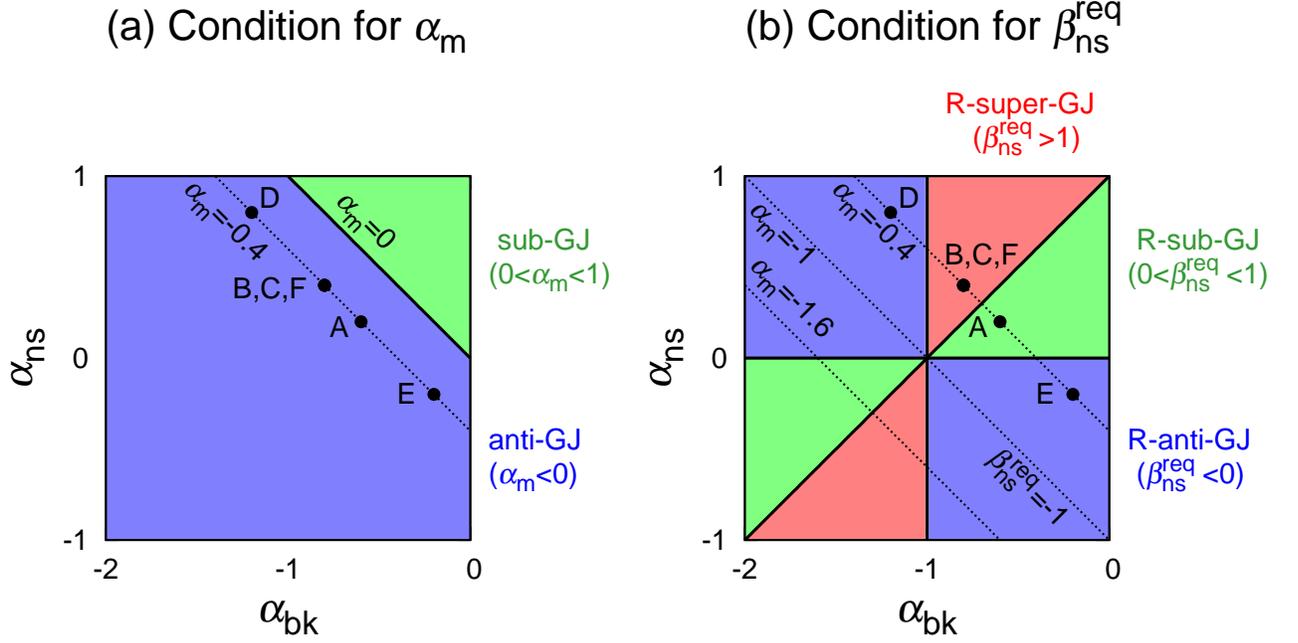}
 \caption{The current density diagrams in the steady-state for $\alpha_{\rm m}$ (panel a) 
and $\beta_{\rm ns}^{\rm req}$ (panel b). 
The red, blue, and green regions correspond to the R-super-GJ (panel b; $\beta_{\rm ns}^{\rm req}>1$), 
anti-GJ (panel a; $\alpha_{\rm m}<0$) or R-anti-GJ (panel b; $\beta_{\rm ns}^{\rm req}<0$), 
and sub-GJ (panel a; $0<\alpha_{\rm m}<1$) or R-sub-GJ (panel b; $0<\beta_{\rm ns}^{\rm req}<1$), respectively. 
The dotted lines show the cases of $\alpha_{\rm m}=\alpha_{\rm ns}+\alpha_{\rm bk}=0$ (a), -0.4 (a and b), -1 (b) and -1.6 (b). 
In panel (b), $\beta_{\rm ns}^{\rm req}=-1, 0$, and 1 correspond to the dotted line $\alpha_{\rm m}=-1$, 
the solid horizontal line $\alpha_{\rm ns}=0$, and the solid diagonal line $\alpha_{\rm ns}=\alpha_{\rm bk}+1$,
respectively.} 
  \label{diagram}
 \end{center}
\end{figure*}

If the condition, 
\begin{equation}
\rho =\rho_{\rm GJ}, ~~
\alpha = \alpha_{\rm m}, ~~
\label{eqn:steadystatesolution}
\end{equation}
is satisfied, 
we have a null solution, namely a steady and an uniform solution with $E_\parallel=0$, where no particle acceleration occurs. 
In the case that particles are supplied only from the NSS, 
equations (\ref{betaByAlphaRhoRhoGJ}) and (\ref{eqn:steadystatesolution}) require 
\begin{eqnarray}\label{eqn:oneflow}
\beta_{\rm ns} = \alpha_{\rm m},
\end{eqnarray}
to screen the electric field $(j=j_{\rm ns})$. 
Then, the condition $0<\alpha_{\rm m}<1$ \citep[hereafter, sub-GJ condition; ][]{S97} 
should be satisfied for the null solution to be possible. 
\footnote{However, even for the sub-GJ case, the null solution is not always guaranteed.
When charged particles are extracted from the NSS with a non-relativistic velocity, 
the velocity increases following the continuity equation, $n\beta=$ constant.
The change of the velocity leads to the spatial change of the charge density, 
which causes the deviation from the GJ density.
Spatially oscillating solutions with averaged electric field $\bar{E}_\parallel$=0 are 
obtained by \citet{Me85} and \citet{S97} for such cases. 
\citet{TA13} and \citet{CB13} pointed out that such oscillating solutions 
are unstable, 
and obtained quasi-steady solutions. 
They observe inefficient particle acceleration and no pair creation with $E_\parallel \sim 0$ 
except for the region just above the NSS. 
Therefore, for the sub-GJ case 
without a back-flow, we can conclude that
particle acceleration is inefficient.}

If the condition $0 < \alpha_{\rm m} < 1$ is not satisfied for the case without back flows, 
particles from the NSS are accelerated and the pair cascade is ignited.
The condition $\alpha_{\rm m}>1$ \citep[hereafter super-GJ condition; ][]{S97} requires superluminal velocity 
for particles from the NSS (equation \ref{eqn:oneflow}), 
so that the velocity $\beta_{\rm ns}$ cannot equal to $\alpha_{\rm m}$ and then the system develops high voltage drops, 
causing $\gamma$-ray emissions due to curvature radiation and intense bursts of pair creation 
\citep{Me85, S97, B08} near the NSS. 
The system with $\alpha_{\rm m} \le 0$ \citep[hereafter anti-GJ condition; ][]{TA13} also develops high voltage.
A recent numerical study \citep{TA13} supports these speculations. 

We focus on only the anti-GJ condition.
Since the current in the OAR should be the return current
($\alpha_{\rm m}<0$; Figure \ref{schematics}) as we mentioned in Section \ref{model},
the current at the inner boundary of the OAR is also expected to be anti-GJ value. 
In this case, when the effect of the back-flowing particles is negligible,
a significant number of pairs are created near the NSS because of the unscreened electric field \citep{TA13}. 
Here, we discuss the possibility that the back-flows can assist the screening of the electric field near the NSS
without pair creation.

In case 1, particles with $\gamma\gg1$ (beam component) flow back to the NSS.
The current densities of the flowing particles from the NSS $j_{\rm ns}$ and the OAR $j_{\rm bk, bm}$ 
are described by dimensionless current densities $\alpha_{\rm ns}$ and $\alpha_{\rm bk, bm}$ as follows:
\begin{equation}
\begin{array}{lcl}
j_{\rm ns} &=&\alpha_{\rm ns}c\rho_{\rm GJ}, \\
j_{\rm bk, bm}&=&\alpha_{\rm bk, bm}c\rho_{\rm GJ}.
\end{array}
\end{equation}
Note that the current density of the beam component $\alpha_{\rm bk, bm}$ is a parameter 
and is determined independently of the global current density parameter $\alpha_{\rm m}$.
We also define the average velocity of the particles from the NSS and OAR,
\begin{eqnarray}\label{eqn:betansbetabk}
\begin{array}{lcl}
\beta_{\rm ns}&=&\displaystyle{ \frac{\alpha_{\rm ns}}{ (\rho_{\rm ns}/\rho_{\rm GJ}) } ,} \\
& & \\
\beta_{\rm bk, bm}&=&\displaystyle{ \frac{\alpha_{\rm bk, bm}}{ (\rho_{\rm bk, bm}/\rho_{\rm GJ}) } ,}\\
\end{array}
\end{eqnarray}
where $\rho_{\rm ns}$ and $\rho_{\rm bk, bm}$ represent the charge densities of the above two components. 
In steady state, 
the sum of the two current densities should satisfy 
\begin{equation}\label{eqn:ainaoutam}
\alpha_{\rm ns} + \alpha_{\rm bk, bm}=\alpha_{\rm m}.
\end{equation}
We only consider the case $\alpha_{\rm bk, bm} < 0$ ($-1 < \beta_{\rm bk, bm} < 0$) and $\alpha_{\rm m} < 0$, 
assuming that the OAR exists on the same field line. 
To screen the electric field uniformly, 
equation (\ref{poisson}) requires $\rho=\rho_{\rm ns}+\rho_{\rm bk, bm}=\rho_{\rm GJ}$, which can be described as,
\begin{equation}
\frac{\alpha_{\rm ns}}{\beta^{\rm req}_{\rm ns}} + \frac{\alpha_{\rm bk, bm}}{\beta_{\rm bk, bm}} = 1.
\end{equation}
From this equation, the required value of $\beta_{\rm ns}$ to screen the field is given by,
\begin{eqnarray}
\beta^{\rm req}_{\rm ns}=\frac{\alpha_{\rm ns}}{1-(\alpha_{\rm bk, bm}/\beta_{\rm bk, bm})},
\label{eqn:EparaScreenCondtion}
\end{eqnarray} 
In case 1, since the current density $j_{\rm bk, bm}$ is carried by ultra-relativistic particles 
($\beta_{\rm bk, bm} \rightarrow -1$), 
equation (\ref{eqn:EparaScreenCondtion}) becomes
\begin{eqnarray}
\beta^{\rm req}_{\rm ns}&=&\frac{\alpha_{\rm ns}}{1+\alpha_{\rm bk, bm}}\nonumber \\
&=&\frac{\alpha_{\rm m}-\alpha_{\rm bk, bm}}{1+\alpha_{\rm bk, bm}}.
\label{eqn:EparaScreenCondtionForRelativisticJout}
\end{eqnarray} 
For the flow from the NSS, we use $\beta_{\rm ns}^{\rm req}$ to characterize the system including the effects of 
the beam component. 
Although the total current density is anti-GJ ($\alpha_{\rm m}<0$), $\beta_{\rm ns}^{\rm req}$ can have arbitrary value.
The cases $\beta_{\rm ns}^{\rm req} < 0$, $0 < \beta_{\rm ns}^{\rm req} < 1$ and $\beta_{\rm ns}^{\rm req} > 1$ 
are similar to the situations of anti-GJ, sub-GJ and super-GJ in the system without the back-flows, respectively. 
Hereafter, those revised versions of the conditions on the current are called such as R-anti-GJ etc., 
respectively.
When the R-sub-GJ condition is satisfied, the electric field can be screened out by the contribution of the back-flow.
In this case, the OAR is steadily maintained with $\alpha_{\rm m}<0$.

Figure \ref{diagram} shows the diagrams for $\alpha_{\rm m}$ (a) and $\beta_{\rm ns}^{\rm req}$ (b) as a function of  
$\alpha_{\rm bk}(=\alpha_{\rm bk, bm}$ in case 1) and $\alpha_{\rm ns}$. 
We focus on the cases where the number and current densities are an order of the GJ value, so that the ranges shown in Figure \ref{diagram} are around $\alpha_{\rm m}\sim-1$ and $\alpha_{\rm bk}\sim-1$.
Since $\alpha_{\rm m}$ in the steady state is the sum of the two current densities (equation \ref{eqn:ainaoutam}), 
the lines corresponding to $\alpha_{\rm m}=$ const. are expressed by the dashed diagonal lines 
from upper left to lower right in this diagram. 
From equation (\ref{eqn:EparaScreenCondtionForRelativisticJout}), 
the lines $\alpha_{\rm ns}=0, -\alpha_{\rm bk}-1, \alpha_{\rm bk}+1$ imply $\beta_{\rm ns}^{\rm req}=0, -1, 1$, respectively. 
For a certain value of $\alpha_{\rm m}$, there are two regions for the R-anti-GJ condition as shown in Figure \ref{diagram} (b)
(blue; $\beta_{\rm ns}^{\rm req}\le-1$, and $-1<\beta_{\rm ns}^{\rm req}\le0$). 
In the red and blue regions in Figure \ref{diagram} (b), the significant pair creation would occur. 

These two diagrams show that 
even if the total current density is anti-GJ, certain ranges of the current density $\alpha_{\rm bk}$ can achieve R-sub-GJ. 
However, the electric field can be screened in 
only some particular combinations of the current densities $\alpha_{\rm bk}$ and $\alpha_{\rm m}$. 
Such an accidental combination may be rarely satisfied.
In order to maintain the OAR steadily, another particle component is required 
unless the combination of $\alpha_{\rm bk, bm}$ and $\alpha_{\rm m}$ are adjusted in the particular range 
by some kinds of the regulation mechanism. 

\section{NUMERICAL SIMULATIONS} 
\label{numerical}

In case 2, the thermal component with relatively small Lorentz factor 
also comes back to the NSS from the OAR [Figure \ref{image} (b)]. 
The thermal component is generally required to screen the electric field at the inner boundary of the OAR as seen 
in the simulations in \citet{T10}. 
In order to treat the particles with finite momentum distribution 
(including non-relativistic cases), we perform the numerical simulations of the encountering plasma just above the NSS 
with the 1D electrostatic Particle-in-Cell code \citep[e.g., ][]{BL85, T10, TA13}. 
First, we confirm the analytical results for case 1 [Figure \ref{image} (a)]
discussed in Section \ref{analytic}.
Then, we consider case 2, where both the beam and thermal components are injected from the outer boundary 
under the anti-GJ condition and investigate how many thermal particles are needed to be injected from the
outer boundary in order to screen the electric field.

\subsection{Numerical Setup} 
\label{setup}

In this subsection, we describe the numerical set-up.
From here, we omit the subscript $\parallel$ from all quantities. 
A coordinate axis $x$ is directed along the field lines, 
its origin is at the NSS and positive direction is towards the OAR. 

In the 1D model the only changing component of electromagnetic fields is the electric field component $E(x,t)$ 
parallel to the x-axis. 
We solve the evolutionary equation for the electric field $E(x,t)$ with,
\begin{eqnarray}\label{ampere2}
\frac{\partial E(x,t)}{\partial t}=-4\pi(j(x,t)-j_{\rm m}).
\end{eqnarray}
Here $j(x,t)$ is the current density at the point $x$ and time $t$.
For the calculation of the current density $j(x,t)$, we use a 1D version of the charge conservative algorithm
proposed by \citet{VB92}.
In order to obtain the initial value of the electric field $E(x,0)$, 
we solve the Poisson equation
\begin{eqnarray}\label{poisson2}
\frac{dE}{dx}(x,0)=4\pi(\rho(x,0)-\rho_{\rm GJ}).
\end{eqnarray}
For the GJ charge density $\rho_{\rm GJ}$, we assume a spatially and temporary constant value in the calculation domain.

To model the free emission of the particles from the NSS ($x=0$), we adopt the same method proposed by \citet{TA13}. 
At the beginning of each time-step, electrons and ions with certain equal number $N_{\rm ns}$ are injected into a cell
just outside the numerical domain $x<0$ (ghost cell) to carry out the electric field calculation.
We adopt a significant number of $N_{\rm ns}$ to correctly simulate the space-charge limitation condition 
\citep[for details, see Appendix C in][]{TA13}.
The momentum of each injected particle is sampled from an uniform distribution in the interval 
$[-p_{\rm ns}, p_{\rm ns}]$ to model non-relativistic and finite temperature of the particles. 
We adopt the momentum $p_{\rm ns}=10^{-2}m_{\rm e}c$, 
though the results do not basically depend on the specific value of $p_{\rm ns}(<m_{\rm e}c)$.
We take the mass ratio of ion to electron as $m_{\rm i}/m_{\rm e}=1836$. 

We introduce $\alpha_{\rm bk, th}$ as the current density of the thermal component. 
Then, the current density of back-flowing particles are described by 
\begin{eqnarray}
\alpha_{\rm bk}=\alpha_{\rm bk, bm}+\alpha_{\rm bk, th}.
\end{eqnarray}
In the steady state, the current densities should satisfy the relation
\begin{eqnarray}\label{eqn:steadytwoflow}
\alpha_{\rm ns}+\alpha_{\rm bk}=\alpha_{\rm m}.
\end{eqnarray}

We inject the back-flowing particles just outside the numerical domain $x>1$ (the last cell) in our simulations, 
where $x$ is normalized by the length of the calculation domain $L$. 
For the beam component, we inject electrons (for $\rho_{\rm GJ}<0$) or positrons (for $\rho_{\rm GJ}>0$) 
with Lorentz factor $\gamma=10^6$ into the calculation domain.
The number density of the injected particles is determined by the parameter $\alpha_{\rm bk, bm}$. 
The particles are injected with a time-step $\delta t$, and the injection point for each particle is 
distributed randomly in space into the last cell. 
The thermal particles from the OAR consist of electrons and positrons with the same number density. 
At the injection point, the average velocity of the thermal component is 
$\bar{\beta}_{\rm bk, th}=0$ and the temperature of them is assumed as $kT_{\rm bk, th}=m_{\rm e}c^2$.
In this set-up, the charge and current densities are $\rho_{\rm bk, th}=0$ and $\alpha_{\rm bk, th}=0$ at the injection point.
The model parameter for the thermal component is only the number density $n_{\rm bk, th}$ in the last cell. 

In the 1D model, 
charged particles are represented by thin sheath with infinite extend in the direction perpendicular to the $x$-axis. 
The equation of motion for a particle $i$ is
\begin{eqnarray}\label{xstep}
\frac{dx_i}{dt}=v_i
\end{eqnarray}
\begin{eqnarray}\label{vstep}
\frac{dp_i}{dt}=\frac{eE(x_i)}{mc}, \ i=1,...,N_{\rm p}
\end{eqnarray}
where $x_i$ and $p_i$ are the position and momentum of the $i$th particle in unit of $\lambda_{\rm pe}$ and $mc$, respectively. 
The length scale $\lambda_{\rm pe}$ is the electron skin depth for the GJ density, 
\begin{eqnarray}
\lambda_{\rm pe}\equiv\frac{c}{\omega_{\rm pe}}=c\biggl(\frac{4\pi|\rho_{\rm GJ}|e}{m_{\rm e}}\biggr)^{-1/2}.
\end{eqnarray}
Since we focus on the nearly screened state of the electric field, particles do not obtain high momentum
(at most $\sim10^6m_{\rm e}c$ even for the beam particles)
in our calculation domain within the reasonable range of $\alpha_{\rm m}$ (an order of unity). 
The radiation drag force starts to work at $\gamma>10^6$ so that we can neglect the reaction force 
in the equation of motion (\ref{vstep}) \citep{T10}.
We have to choose the total number of the simulation particles $N_{\rm p}$ and 
the length of the calculation domain $L$ in order to satisfy the conditions
\begin{eqnarray}\label{NL}
L\gg\lambda_{\rm pe},\ N_{\rm p}\gg\frac{L}{\lambda_{\rm pe}}.
\end{eqnarray}
In this limit, the results are expected to be independent of the choice of $N_{\rm p}$ and $L$. 
In our simulations, typically $L=(10^2$-$10^3)\lambda_{\rm pe}$ and $N_{\rm p}\sim10^6$. 
Another requirement is a significantly short time-step in the simulation, $\delta t\ll\omega_{\rm pe}$, 
to resolve plasma oscillations. 
In our simulations, non-relativistic particles play an important role to screen the electric field. 
Our choices for the time step $\delta t$ and each grid size $\delta x$
are always $\delta t = 0.02\omega_{\rm pe}$ and $\delta x = 0.1\lambda_{\rm pe}$.

Initially, we set a condition that there is no particle in the calculation domain ($\rho(x,0)=0$). 
We have checked that the results basically do not depend on the initial particle distribution 
once the system reaches the quasi-steady state. 

The extracted particles (electrons or ions) from the NSS in the quasi-steady state 
depend on the signs of the parameters, $\alpha_{\rm m}-\alpha_{\rm bk, bm}$ and $\rho_{\rm GJ}$ 
\footnote{Note that before reaching the quasi-steady state, both particles can be extracted from the NSS 
depending on the electric field just above the NSS.}. 
Equation (\ref{eqn:ainaoutam}) indicates that 
the signs of $\alpha_{\rm ns}$ and $\alpha_{\rm m}-\alpha_{\rm bk,bm}$ are the same in the quasi-steady state. 
Since $\beta_{\rm ns}$ is always positive, the sign of $\rho_{\rm ns}$ is the same as (opposite to) that of 
$\rho_{\rm GJ}$ for positive (negative) $\alpha_{\rm ns}$ (see equation \ref{eqn:betansbetabk}) 
or equivalently positive (negative) $\alpha_{\rm m}-\alpha_{\rm bk, bm}$.
The parameter sets we adopt are summarized in Table 1. 

Although we only consider $\alpha_{\rm m}=-0.4$ in Models A-F, 
the following results are the same as the case $\alpha_{\rm m}=-1.6$. 
The diagram in Figure \ref{diagram} (b) is symmetrical to the point $(\alpha_{\rm bk}, \alpha_{\rm ns})=(-1,0)$. 
For example, electrons with the average density $\bar{n}_{\rm ns}/n_{\rm GJ}=0.4$ and velocity $\bar{\beta}_{\rm ns}=0.5$ 
are extracted from the NSS in Model A. 
In the anti-pulsar case ($\rho_{\rm GJ}>0$) with the current densities $\alpha_{\rm m}=-1.6$ and $\alpha_{\rm bk, bm}=-1.4$, 
the electrons with average density $\bar{n}_{\rm ns}/n_{\rm GJ}=0.4$ and velocity $\bar{\beta}_{\rm ns}=0.5$ are also extracted from the NSS 
to screen the electric field from equation (\ref{eqn:EparaScreenCondtionForRelativisticJout}).
Then, the two cases, $(\alpha_{\rm m}, \alpha_{\rm bk, bm}, \rho_{\rm GJ})$ 
and $(\alpha_{\rm m}', \alpha_{\rm bk, bm}', \rho_{\rm GJ}')=(-\alpha_{\rm m}-2, -\alpha_{\rm bk,bm}-2, -\rho_{\rm GJ})$, 
result in the same solution for the particle flow from the NSS.

\begin{table*}
\begin{center}
\begin{tabular}{lcccccc}
\multicolumn{7}{c}{TABLE 1 Simulation model parameters} \\ \hline
Model & $\rho_{\rm GJ}$ & $\alpha_{\rm m}$ & $\alpha_{\rm bk, bm}$ & $\beta_{\rm ns}^{\rm req}$ &  & $n_{\rm bk, th}/n_{\rm GJ}$  \\ \hline
A    & negative & $-0.4$          & $-0.6$            & $0.5$      & R-sub-GJ        & $0$                    \\
B    & negative & $-0.4$          & $-0.8$            & $2.0$      & R-super-GJ      & $0$                    \\
C1   & negative & $-0.4$          & $-0.8$            & $2.0$      & R-super-GJ      & $0.1$              \\
C2   & negative & $-0.4$          & $-0.8$            & $2.0$      & R-super-GJ      & $0.2$              \\
C3   & negative & $-0.4$          & $-0.8$            & $2.0$      & R-super-GJ      & $0.4$              \\
C4   & negative & $-0.4$          & $-0.8$            & $2.0$      & R-super-GJ      & $0.6$              \\
C5   & negative & $-0.4$          & $-0.8$            & $2.0$      & R-super-GJ      & $0.8$              \\
C6   & negative & $-0.4$          & $-0.8$            & $2.0$      & R-super-GJ      & $1.0$              \\
D1   & negative & $-0.4$          & $-1.2$            & $-4.0$     & R-anti-GJ       & $0.2$               \\
D2   & negative & $-0.4$          & $-1.2$            & $-4.0$     & R-anti-GJ       & $0.6$               \\
D3   & negative & $-0.4$          & $-1.2$            & $-4.0$     & R-anti-GJ       & $1.0$               \\
D4   & negative & $-0.4$          & $-1.2$            & $-4.0$     & R-anti-GJ       & $1.4$               \\
D5   & negative & $-0.4$          & $-1.2$            & $-4.0$     & R-anti-GJ       & $1.8$               \\
D6   & negative & $-0.4$          & $-1.2$            & $-4.0$     & R-anti-GJ       & $2.2$               \\
E1   & positive & $-0.4$          & $-0.2$            & $-0.25$    & R-anti-GJ       & $0.5$               \\
E2   & positive & $-0.4$          & $-0.2$            & $-0.25$    & R-anti-GJ       & $0.6$               \\
E3   & positive & $-0.4$          & $-0.2$            & $-0.25$    & R-anti-GJ       & $0.7$               \\
E4   & positive & $-0.4$          & $-0.2$            & $-0.25$    & R-anti-GJ       & $0.8$               \\
E5   & positive & $-0.4$          & $-0.2$            & $-0.25$    & R-anti-GJ       & $0.9$               \\
E6   & positive & $-0.4$          & $-0.2$            & $-0.25$    & R-anti-GJ       & $1.0$               \\
F1   & positive & $-0.4$          & $-0.8$            & $2.0$      & R-super-GJ      & $0.2$              \\ 
F2(a-e) & positive & $-0.4$       & $-0.8$            & $2.0$      & R-super-GJ      & $0.4$              \\ 
F3   & positive & $-0.4$          & $-0.8$            & $2.0$      & R-super-GJ      & $0.6$              \\ 
F4   & positive & $-0.4$          & $-0.8$            & $2.0$      & R-super-GJ      & $1.1$              \\ 
F5   & positive & $-0.4$          & $-0.8$            & $2.0$      & R-super-GJ      & $1.6$              \\ \hline
\multicolumn{7}{l}{}%
\label{tab:parameter}
\end{tabular}
\end{center}
\end{table*}

\subsection{Single Beam Back-flow}
\label{single}

\begin{figure*}
 \begin{center}
  \includegraphics[width=120mm, angle=270]{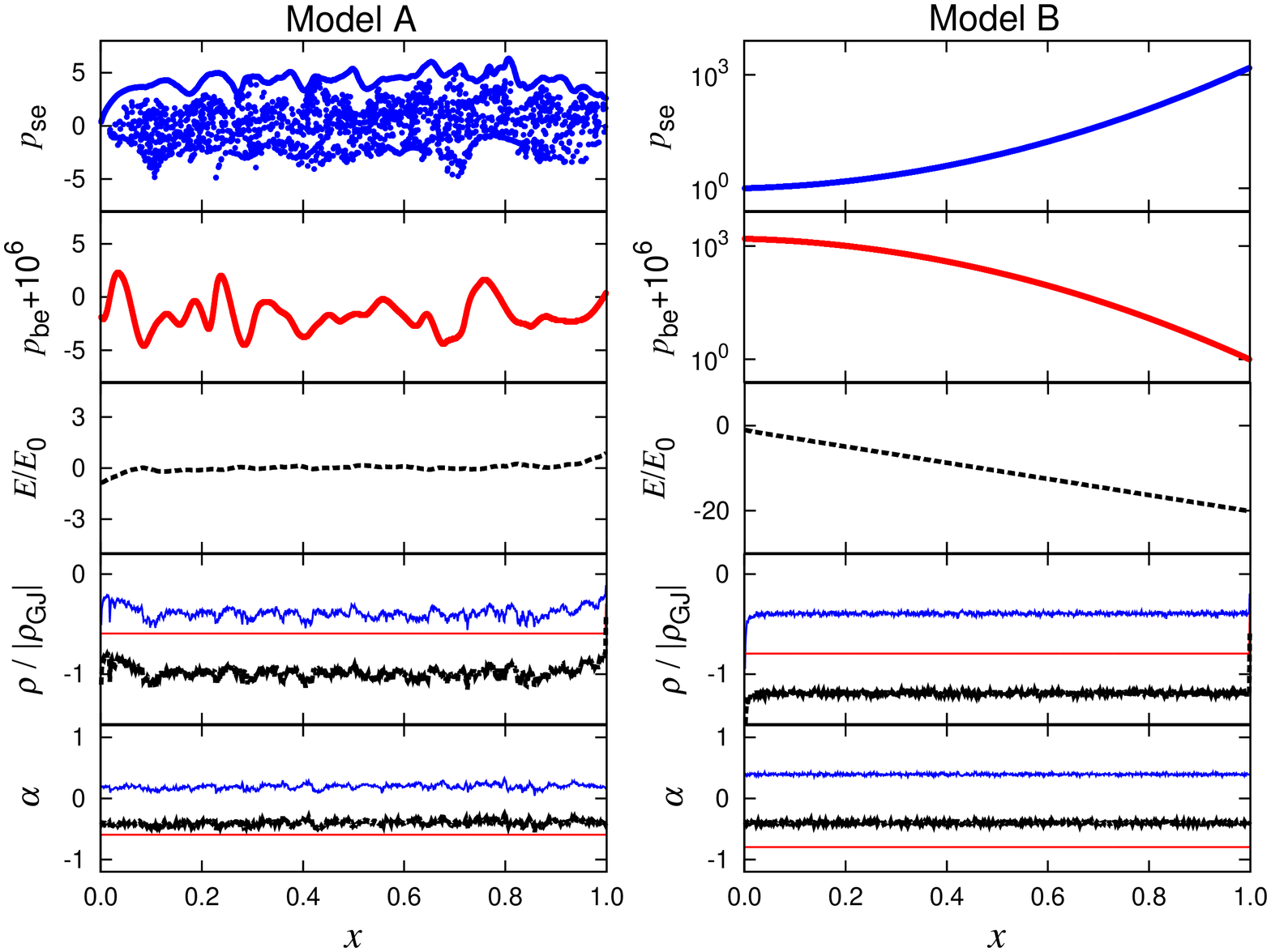}
 \end{center}
 \caption{Snapshots of the quasi-steady state for Models A (R-sub-GJ, left) and B (R-super-GJ, right). 
Each column gives the following physical quantities in the computational domain: 
momentum (normalized by $m_{\rm e}c$) of electrons from the NSS $p_{\rm se}$ (blue), that from the OAR $p_{\rm be}+10^6$ (red),  
accelerating electric field $E$ normalized by $E_0\equiv m_{\rm e}c\omega_{\rm pe}/e$, charge densities $\rho/|\rho_{\rm GJ}|$, 
and current densities $\alpha$. 
The length of the calculation domain (from $x=0$ to $x=1$) is $L=100\lambda_{\rm pe}$. 
For the charge and current densities (fourth and fifth columns), black lines denote the total values. 
Blue and red lines imply the contributions of particles from the NSS and the outer boundary, 
respectively. }
 \label{sub1}
\end{figure*}

In the left panels of Figure \ref{sub1}, we show a snapshot for different physical quantities
in quasi-steady state of Model A ($\alpha_{\rm m}=-0.4, \alpha_{\rm bk, bm}=-0.6$),
for which $\beta_{\rm ns}^{\rm req}=0.5$ satisfies the R-sub-GJ condition.
In the case $\beta_{\rm ns}^{\rm req}=0.5$,
the extracted electrons may behave similarly to the sub-GJ case of $\alpha_{\rm m}=0.5$
without the beam component \citep{TA13, CB13}. 
The figure shows that electric field is screened in most of the calculation domain as we expected. 
In the momentum distribution of the extracted electrons, two components, 
the cold outgoing beam and trapped thermal ones, are seen.
The cold beam particles from the NSS adjusts the total current density to $\alpha_{\rm m}$. 
Because particles from the NSS initially have non-relativistic velocity, 
the absolute value of the normalized charge density $|\rho/\rho_{\rm GJ}|$ just above the surface is larger than unity 
in the steady state. 
Then, the non-relativistic particles are accelerated by the generating electric field 
and the normalized charge density $|\rho/\rho_{\rm GJ}|$ decreases toward the OAR. 
The accelerated particles mainly compose of the outgoing beam component. 
The trapped component in the calculation domain adjusts the total charge density to $\rho_{\rm GJ}$ 
except for the region just above the NSS. 
The formation of this trapped thermal component is attributed to the instability 
as discussed by previous authors \citep{TA13, CB13}, 
who show that the spatially oscillated solution in cold limit on the particle momentum \citep[e.g., ][]{S91} 
is unstable for the particles with non-negligible temperature at the NSS.
The trapped particles are reflected by the electric field near the boundaries, so that most of them cannot take part in 
the outgoing beam component.
The electric-field structure within $0<x<0.1$ can be regarded as the relativistic double layer, 
which is defined as a large potential drop ($\gg m_{\rm e}c^2$) maintained by an anode and a cathode 
neglecting gravity and thermal plasma motion at the boundaries \citep[e.g., ][]{C82}. 
The electric field also appears near the outer boundary ($0.9<x<1.0$) to trap the thermal particles. 
The results for the extracted particles from the NSS 
are close to the sub-GJ case without back-flowing component \citep{TA13, CB13}, 
so that even if $\alpha_{\rm m}< 0$, the electric field is screened and no significant particle acceleration occurs 
when $0 < \beta_{\rm ns}^{\rm req}<1$. 
The tiny momentum fluctuation of the beam component from the outer boundary 
(seen in the left second column of Figure \ref{sub1}) is due to the fluctuation in the charge density of the trapped particles. 
For the beam component from the OAR, the velocity ($\sim-c$) and the current density are almost constant as shown 
in the left panels of Figure \ref{sub1}.

In the right panels of Figure \ref{sub1}, we also show the quasi-steady state for the R-super-GJ case (Model B). 
We set the sign of $\rho_{\rm GJ}$ is positive and inject positrons as beam particles from the outer boundary.
Then, mainly extracted particles from the NSS are electrons in Model B. 
Obviously, particles from both boundaries are continuously accelerated. 
Because our calculation domain is not so large, the change of Lorentz factor is $\sim 10^3$. 
In reality, the value of the Lorentz factor would reach the acceleration-reaction limit. 
As a result, electrons and positrons emit $\gamma$-ray photons and significant pair creation should occur. 
Therefore, we confirm that the beam component from the OAR does not assist the screening for the R-super-GJ case,
as mentioned in Section \ref{analytic}. 
In order to screen the field without pair creation in the anti-GJ case, 
the R-sub-GJ condition is required, as long as only an ultra-relativistic component is considered as the back-flow.

\subsection{Beam and Thermal Back-flows}
\label{twoflow}

In this subsection, we assume that both the beam and thermal components come back to the NSS 
under the anti-GJ condition ($\alpha_{\rm m}=-0.4$). 
The thermal particles may coherently interact with the particles extracted from the NSS. 
Then, the results depend on the relative mass of the particles. 
We separately consider the electron-extracted case (Section \ref{electron}) and the ion-extracted case (Section \ref{ion})
from the NSS.
Hereafter, we classify cases as R-anti-GJ, etc. 
based on equation (\ref{eqn:EparaScreenCondtionForRelativisticJout}).
As we will see later, if the number density of thermal particles is large enough, 
their contribution finally changes the status of the whole calculation domain to R-sub-GJ.

\subsubsection{Electron-extracted case}
\label{electron}

For a given parameter $\alpha_{\rm m}$, a dotted line in Figure \ref{diagram} (b) shows that 
the system can be the R-super-GJ, R-sub-GJ, and two R-anti-GJ regions 
($\alpha_{\rm bk}<-1, \alpha_{\rm bk}>-1$) for different $\alpha_{\rm bk}$. 
We investigate the screening conditions for the R-super-GJ condition
(Models C1-6; $\beta_{\rm ns}^{\rm req}=2.0$)
and for the two kinds of the R-anti-GJ conditions
(Models D1-6; $\beta_{\rm ns}^{\rm req}=-4.0$, and Models E1-6; $\beta_{\rm ns}^{\rm req}=-0.25$). 
The length of the calculation domain is $L=100\lambda_{\rm pe}$.
Since the beam back-flow is not significantly decelerated in our results, the charge and current densities
of the beam component are constant in the calculation domain as seen in Section \ref{single}. 
We omit showing the momentum portrait for the beam component from the OAR in Section \ref{twoflow}. 

\begin{figure*}
 \vspace{-2.5cm}
 \begin{center}
  \includegraphics[width=160mm, angle=270]{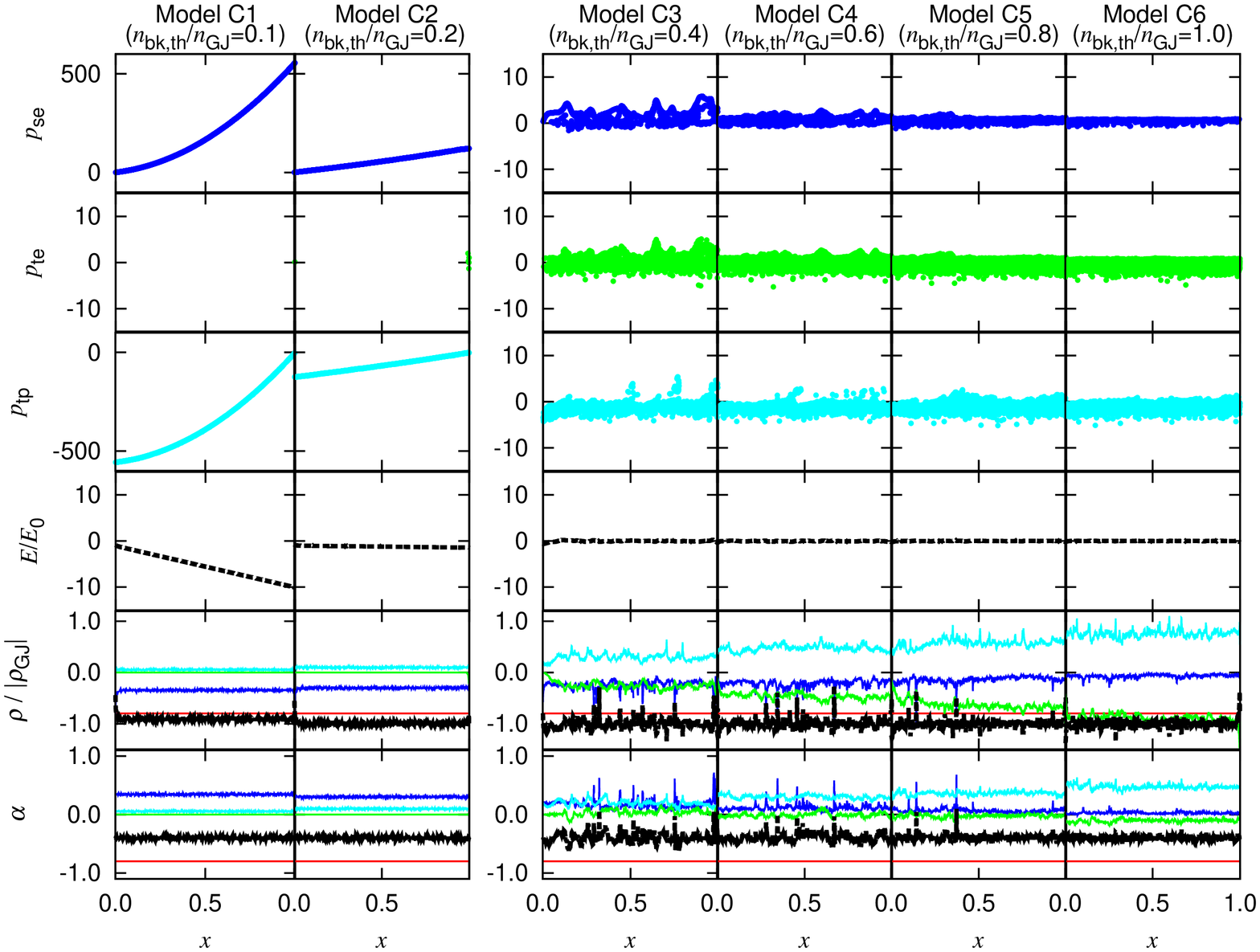}
 \caption{Snapshots of the quasi-steady state for Models C1-6 (starting from R-super-GJ, $\beta_{\rm ns}^{\rm req}=2.0$). 
From left to right rows, the injected number densities of the thermal electrons and positrons from the outer boundary
$n_{\rm bk, th}/n_{\rm GJ}$ are 0.1 (Model C1), 0.2 (Model C2), 0.4 (Model C3), 0.6 (Model C4), 0.8 (Model C5) and 1.0 (Model C6). 
The first (momentum of electrons from the NSS), forth (the electric field), fifth (charge densities), 
and sixth columns (current densities) 
are the same plots as in Figure \ref{sub1}. 
The length of the calculation domain is $L=100\lambda_{\rm pe}$.
The second and third columns show the normalized momenta of the thermal electrons 
$p_{\rm te}$ (green) and positrons $p_{\rm tp}$ (cyan) coming from the OAR. 
The contributions of the thermal particles from the OAR are shown 
as the same colors in the fifth and sixth columns. 
The red lines in the fifth and sixth columns show the contribution of the beam electrons from the OAR with $\gamma\sim10^6$. 
For the first and third columns in Models C1 and C2, 
we show the wide range of momenta, $-100\le p_{\rm se}\le 600$ and $-600\le p_{\rm tp}\le100$, respectively.
In the other models the ranges are taken as $-15\le p_{\rm se}, p_{\rm te}, p_{\rm tp}\le15$. }
 \label{super05}
 \end{center}
\end{figure*}
\begin{figure}
 \begin{center}
 \hspace{-15mm}
  \includegraphics[width=70mm, angle=270]{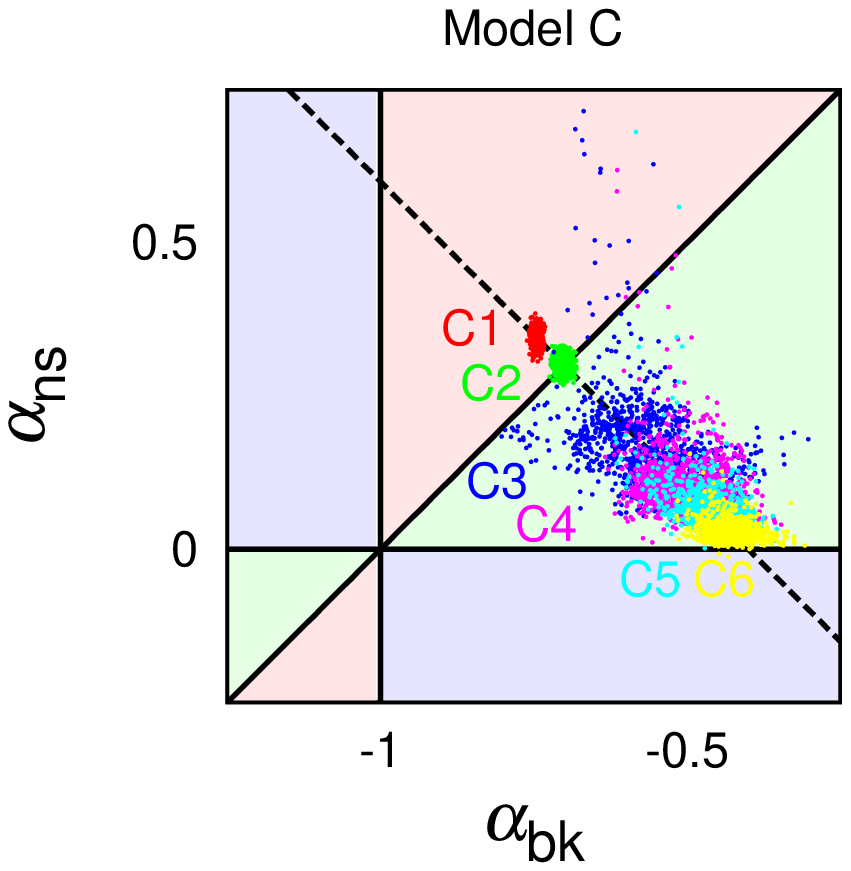}
 \caption{Snapshots of the current densities, $\alpha_{\rm bk}=\alpha_{\rm bk, bm}+\alpha_{\rm bk, th}$ and $\alpha_{\rm ns}$, 
in quasi-steady state of Models C1-6 for each point $x$. 
The meanings of the solid lines and the colors of the areas are the same as in Figure \ref{diagram} (b). 
The dashed line shows $\alpha_{\rm ns}+\alpha_{\rm bk}=\alpha_{\rm m}=-0.4$. 
The red, green, blue, magenta, cyan and yellow points show the cases of 
$n_{\rm bk, th}/n_{\rm GJ}=$0.1 (Model C1), 0.2 (Model C2), 0.4 (Model C3), 0.6 (Model C4), 0.8 (Model C5) and 1.0 (Model C6), 
respectively. } 
 \label{super06}
 \end{center}
\end{figure}

In Figure \ref{super05}, we present the snapshots of the quasi-steady state starting from the R-super-GJ condition
(Models C1-6; $\alpha_{\rm bk, bm}=-0.8$ and $\beta_{\rm ns}^{\rm req}=2.0$) 
for different number densities of the thermal electrons and positrons at the outer boundary. 
The parameters $\alpha_{\rm m}$ and $\alpha_{\rm bk,bm}$ are the same as in Model B.
The sign of $\rho_{\rm GJ}$ is negative, so that the beam particles from the OAR are electrons.
In Model C1 $(n_{\rm bk, th}/n_{\rm GJ}=0.1$; the first row in Figure \ref{super05}), the electric field is not screened. 
The value of the current density $-0.7<\alpha_{\rm bk}<-0.4$ is required to achieve the R-sub-GJ [Figure \ref{diagram} (b)].
Even if all the thermal positrons initially have a velocity $\beta_{\rm bk,th}\sim-1$ and act like the beam particles,
the total current density of the flow from the OAR become
$\alpha_{\rm bk}=\alpha_{\rm bk, bm}+en_{\rm bk, th}\times(-c)/(\rho_{\rm GJ}c)=-0.7$ 
($\beta_{\rm ns}^{\rm req}=1$), which is just at the boundary of the required condition. 
Considering the slow average velocity $|\bar{\beta}_{\rm bk,th}|\ll1$ at the outer boundary, 
the current density becomes $\alpha_{\rm bk}<-0.7$ due to the continuity equation.
Furthermore if the electric field is screened at the outer boundary, 
only a half number density of the thermal positrons $0.5n_{\rm bk, th}$ with an initial momentum $p_{\rm tp}<0$
can enter the calculation domain. 
Thus, the injected number density of the thermal particles $n_{\rm bk, th}/n_{\rm GJ}=0.1$
is too small to steadily screen the electric field. 
In order to clarify this result, 
we plot the value of the current densities $\alpha_{\rm bk}=\alpha_{\rm bk, bm}+\alpha_{\rm bk, th}$ and $\alpha_{\rm ns}$
in Figure \ref{super06} 
for each position $x$ in the calculation domain on the same diagram as Figure \ref{diagram} (b). 
The dashed line denotes the condition $\alpha_{\rm ns}+\alpha_{\rm bk}=\alpha_{\rm m}=-0.4$. 
In Model C1 (red dots), 
the dots still locate the R-super-GJ region (red region). 
When the currents achieve the quasi-steady state, the dots should be along the dashed line. 
Actually the results show $\alpha_{\rm bk}+\alpha_{\rm ns}=\alpha_{\rm m}$ on average 
and the system can be considered as the quasi-steady state 
($\partial E/\partial t\sim0$ from equation \ref{ampere2}). 

In Model C2 $(n_{\rm bk, th}/n_{\rm GJ}=0.2$; the second row in Figure \ref{super05}), 
the accelerating electric field still exists over the calculation domain.
Because of the electric field near the outer boundary, the back-flowing thermal positrons 
are accelerated toward the NSS 
(the third column in Figure \ref{super05}).
Most electrons of the thermal component do not contribute to the current density in the calculation domain 
(the second column in Figure \ref{super05}). 
As shown in Figure \ref{super06},
the center of the distribution of the current densities (green dots) 
locates $\beta_{\rm ns}^{\rm req}=1$ (solid diagonal line). 
Since particles cannot reach $|v|=c$ strictly, 
this solution still requires the electrons from the NSS to be accelerated continuously
\citep[e.g., ][]{M74}. 

When $n_{\rm bk, th}/n_{\rm GJ}>0.2$,
the electric field is almost screened over the calculation domain 
except for the regions near the inner and outer boundaries. 
In Figure \ref{super05}, 
no significant particle acceleration is seen in Models C3-6 ($n_{\rm bk, th}/n_{\rm GJ}=0.4, 0.6, 0.8$ and 1.0, respectively). 
The momentum distributions of the thermal component from the OAR 
are mainly determined by the initial temperature $kT_{\rm bk,th}=m_{\rm e}c^2$. 
Note that in contrast to the particles from the NSS, 
both thermal electrons and positrons from the OAR enter the calculation domain. 
In Model C3 (the third row in Figure \ref{super05}),
although the electric field near the outer boundary accelerates positrons up to $\sim-c$, 
the momentum of the accelerated particle is an order of $\sim -m_{\rm e}c$ in the steady state.
Since the initial temperature at the outer boundary is comparable to the electron rest-mass energy, 
a part of them has a mildly relativistic velocity $\gamma\sim3-5$.
Such mildly relativistic electrons are not reflected by the electric field
near the outer boundary and penetrate the calculation domain. 
On the other hand, at the inner boundary, injected particles have non-relativistic velocity 
so that only electrons accelerated by the electric field near the inner boundary can enter the calculation domain.

In Models C5 and C6, 
the thermal particles from the OAR largely contribute to the total charge and current densities 
(the fifth and sixth columns of Figure \ref{super05}).
These contributions adjust the the total current density to the stationary condition $\alpha=\alpha_{\rm m}$. 
As shown in Figure \ref{super06}, 
almost all dots for $n_{\rm bk,th}/n_{\rm GJ}\ge 0.4$
locate the R-sub-GJ region except for a small number of dots in the R-super-GJ region. 
The electrons from the NSS contribute to the total current density 
($\alpha_{\rm ns}\neq0$; see blue lines in the fifth column of Figure \ref{super05}). 
Such electrons have to be accelerated to significantly contribute to the total current density.
The electric-field structure near the NSS can be regarded as the relativistic double layer, 
which is similar to the result in R-sub-GJ case without the thermal components from the OAR (Figure \ref{sub1}). 
The dispersion of the dots in the R-sub-GJ region in Figure \ref{super06} 
is caused by the plasma oscillation of trapped particles in the calculation domain, 
which adjust the charge density to the GJ value.
The contribution of $\alpha_{\rm ns}$ to the total current density becomes small 
as $n_{\rm bk, th}$ increases. 
Especially for Model C6, 
$\alpha_{\rm ns}\sim0$ in the calculation domain.
In this case, only the thermal component from the OAR adjusts both the current and charge densities 
and the structure of the double layer near the NSS is no longer sustained. 
We also perform the simulations with a larger injection number density $n_{\rm bk, th}/n_{\rm GJ}>1.0$ 
and confirm that only the thermal component adjusts the current and charge densities to 
$\alpha=\alpha_{\rm bk}=\alpha_{\rm m}$ and $\rho=\rho_{\rm GJ}$ in the R-sub-GJ region. 
This is because the thermal electrons from the OAR screen 
the electric field instead of the electrons from the NSS. 
This practically indicates elimination of the polar cap in the anti-GJ region;
the entire region except for the OAR is steadily filled with a high-density plasma
that screens the electric field.

\begin{figure}
 \begin{center}
 \hspace{-15mm}
  \includegraphics[width=70mm, angle=270]{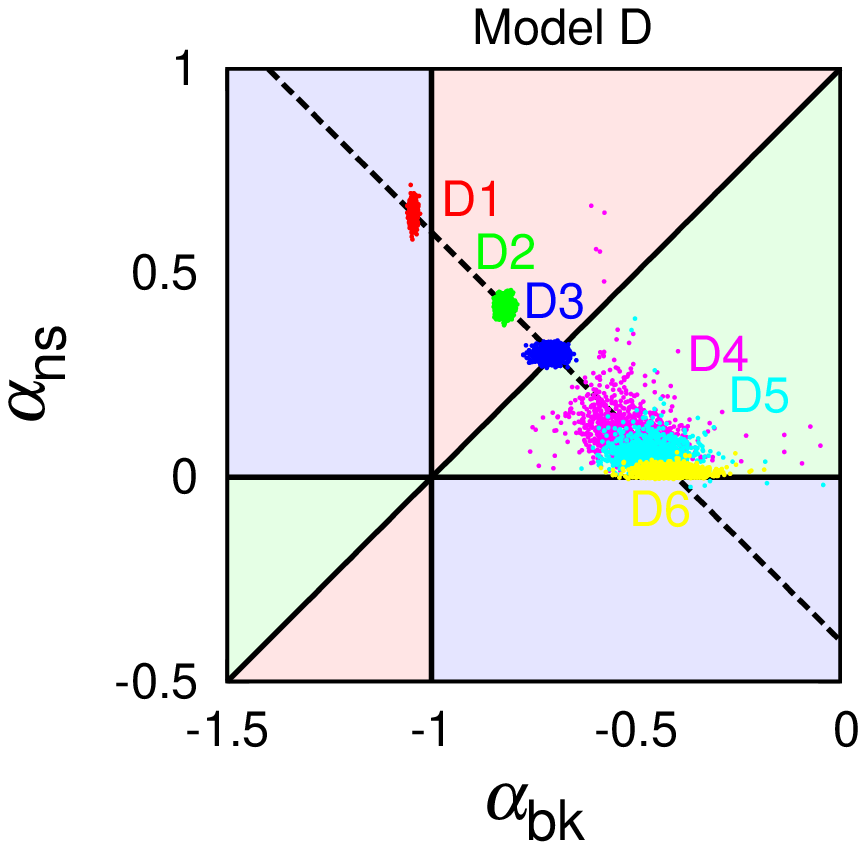}
 \caption{Same as Figure \ref{super06} but for Models D1-6
(starting from the R-anti-GJ, $\beta_{\rm ns}^{\rm req}=-4.0$). 
   The cases of $n_{\rm bk, th}/n_{\rm GJ}=$ 0.2 (red; Model D1), 0.6 (green; Model D2),
   1.0 (blue; Model D3), 1.4 (magenta; Model D4), 1.8 (cyan; Model D5)  and 2.2 (yellow; Model D6) are shown.} 
 \label{anti11}
 \end{center}
\end{figure}

Let us move on to the case starting from the R-anti-GJ condition. 
First, we investigate the case $\beta_{\rm ns}^{\rm req}\le-1$ (Models D1-6; 
$\alpha_{\rm m}=-0.4$, $\alpha_{\rm bk, bm}=-1.2$, so that $\beta_{\rm ns}^{\rm req}=-4.0$).
In Figure \ref{anti11}, 
we present the current densities ($\alpha_{\rm bk}, \alpha_{\rm ns}$) at each position in the quasi-steady state 
for Models D1-6. 
Similarly to the results in the case starting from the R-super-GJ condition (Models C1-6), 
the distribution of dots shift toward right side along the line of
$\alpha_{\rm ns}+\alpha_{\rm bk}=\alpha_{\rm m}$ 
as $n_{\rm bk,th}$ increases. 
In Model D1 $(n_{\rm bk, th}/n_{\rm GJ}=0.2$; red dots in Figure \ref{anti11}),
the current density is $\alpha_{\rm bk}\sim-1.0$. 
This means that almost all injected thermal positrons are accelerated up to $\sim-c$ 
due to the large electric field near the outer boundary.
Because the average velocity of particles which enter the calculation domain is 
$|\bar{\beta}_{\rm bk,th}|\ll 1$ at the outer boundary and the continuity equation is satisfied, 
the current density $\alpha_{\rm bk}$ is a bit smaller than $-1.0$. 
As the density $n_{\rm bk, th}$ increases, 
the contribution of $\alpha_{\rm ns}$ decreases. 
The electric field is almost screened out, when the dots are on the green region in Figure \ref{anti11}.
Those results are almost the same as the case starting from the R-super-GJ condition (Models C1-6; Figure \ref{super05}).

\begin{figure}
 \begin{center}
 \hspace{-15mm}
  \includegraphics[width=70mm, angle=270]{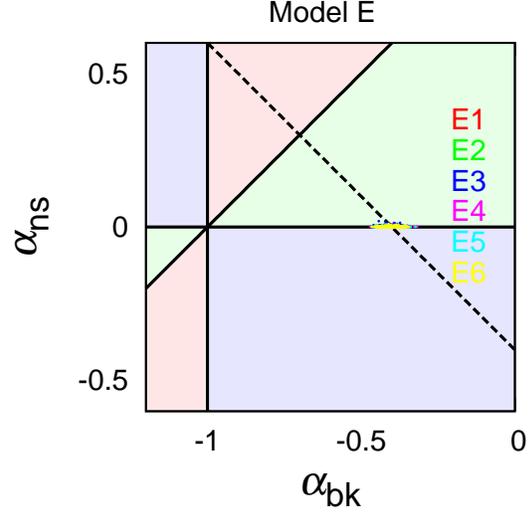}
 \caption{Same as Figure \ref{super06} but for Models E1-6 
(starting from the R-anti-GJ, $\beta_{\rm ns}^{\rm req}=-0.25$).
The cases of $n_{\rm bk, th}/n_{\rm GJ}=$0.5 (red; Model E1), 0.6 (green; Model E2), 
0.7 (blue; Model E3), 0.8 (magenta; Model E4), 
0.9 (cyan; Model E5) and 1.0 (yellow; Model E6). 
Note that almost all the data points are overlapped in the diagram.} 
 \label{anti11_2}
 \end{center}
\end{figure}

Next, we consider another R-anti-GJ condition
(Models E1-6; $\alpha_{\rm bk, bm}=-0.2$, $\alpha_{\rm m}=-0.4$ and $\beta_{\rm ns}^{\rm req}=-0.25$), 
in which the sign of $\alpha_{\rm m}-\alpha_{\rm bk, bm}$ is negative.
To investigate the electron-extracted case,
we assume that the sign of $\rho_{\rm GJ}$ is positive and the beam particles from the OAR are positrons. 
The thermal component from the OAR succeeds to decrease the current density $\alpha_{\rm bk}$ down to $-0.4$.
We show the results in the current density diagram of Figure \ref{anti11_2}. 
All the cases for different values of $n_{\rm bk, th}$ enter the R-sub-GJ region in the diagram,
and the electric field is screened in the calculation domain.
In the cases with $\beta_{\rm ns}^{\rm req}<-1$, 
there are steady-state solutions with $\alpha_{\rm ns}\neq0$ in the R-sub-GJ region (Models D4 and D5).
However, in the case $-1<\beta_{\rm ns}^{\rm req}<0$ (Models E1-6), the current density $\alpha_{\rm ns}$ is always $\sim0$ 
in the screened solutions.
This tendency is irrespective of whether the extracted particles are ions or electrons. 

Our numerical results shown above provide the condition to screen the electric field. 
In the electron-extracted case, we divide non-R-sub-GJ regions into $|\beta_{\rm ns}^{\rm req}|\ge1$
and $-1<\beta_{\rm ns}^{\rm req}<0$. 
In the former case, 
the injection of the thermal component tends to reduce $|\alpha_{\rm ns}|$ 
keeping $\alpha_{\rm ns}+\alpha_{\rm bk}=\alpha_{\rm m}$. 
If we inject the thermal component enough to achieve 
$\beta_{\rm ns}^{\rm req}<1$, 
the electric field is screened over the calculation domain. 
When the system enter the R-sub-GJ region with $\alpha_{\rm ns}\neq0$, 
particles are extracted from the NSS. 
In order to derive the screening condition, 
we introduce a parameter, $n_0$, 
which is the density of particles entering the calculation domain without the electric field 
at the outer boundary 
(i.e., $n_0=0.5n_{\rm bk, th}$ for $\bar{\beta}_{\rm bk, th}=0$). 
If electrons or positrons of the thermal component are accelerated up to $\sim-c$ in the calculation domain,
the density $n_0$ is described as $n_0/n_{\rm GJ}=|\alpha_{\rm bk}-\alpha_{\rm bk, bm}|$. 
From equation (\ref{eqn:EparaScreenCondtionForRelativisticJout}), the screening condition is 
\begin{eqnarray}
\frac{\alpha_{\rm m}-\alpha_{\rm bk}}{1+\alpha_{\rm bk}}<1.
\end{eqnarray}
Then, we derive the required density of the thermal component $n_0$ as,
\begin{eqnarray}\label{screen1}
\frac{n_0}{n_{\rm GJ}}>\left\{\begin{array}{ll}
\frac{1}{2}|(\alpha_{\rm bk, bm}+1)(\beta_{\rm ns}^{\rm req}-1)| & ~{\rm for}~~\alpha_{\rm bk, bm}\neq -1, \\
 & \\
\frac{1}{2}|\alpha_{\rm m}+1| & ~{\rm for}~~\alpha_{\rm bk, bm}=-1. \\
\end{array} \right.
\end{eqnarray}
Note that $\beta_{\rm ns}^{\rm req}$ is defined in equation (\ref{eqn:EparaScreenCondtionForRelativisticJout}) 
neglecting $\alpha_{\rm bk,th}$.
On the other hand, for the case $-1<\beta_{\rm ns}^{\rm req}\le0$, 
the screened state always implies $\alpha_{\rm ns}=0$ ($\alpha_{\rm m}=\alpha_{\rm bk}$).
From equation (\ref{eqn:EparaScreenCondtionForRelativisticJout}), the screening condition is 
\begin{eqnarray}
\alpha_{\rm m}>\alpha_{\rm bk},
\end{eqnarray}
so that we derive the required density $n_0$ to screen the electric field as, 
\begin{eqnarray}\label{screen2}
n_0>|\beta_{\rm ns}^{\rm req}(\alpha_{\rm bk, bm}+1)|.
\end{eqnarray}
Note that this equation also has a singular point at $\alpha_{\rm bk, bm}=-1$. 
However, $\alpha_{\rm bk, bm}=-1$ corresponds to $\beta_{\rm ns}^{\rm req}=\pm\infty$, 
which is not included in the case $-1<\beta_{\rm ns}^{\rm req}\le0$. 
The results do not depend on $L$ and $N_{\rm p}$ if the conditions (\ref{NL}) are satisfied. 

\subsubsection{Ion-extracted case}
\label{ion}

Here, we investigate the case in which extracted particles from the NSS are ions. 
We start from the R-super-GJ condition, in which the parameters of the current density are the same as Models B and C1-6 
(Models F1-5; $\alpha_{\rm bk, bm}=-0.8$ and $\alpha_{\rm m}=-0.4$). 
To consider the ion-extracted case, we assume that the GJ charge density $\rho_{\rm GJ}$ is positive.
Then, the beam particles from the OAR are positrons.
In the electron-extracted cases (Section \ref{electron}), we set the length of the calculation domain $L=100\lambda_{\rm pe}$.
Because the ion skin depth is large, $\lambda_{\rm pp}\sim43\lambda_{\rm pe}$, 
we have to take a larger calculation domain to correctly simulate the dynamics including electrons and ions. 
According to the limited calculation time, we set $L=10^3\lambda_{\rm pe}$ 
and the grid size is the same as the previous calculations.

\begin{figure*}
 \begin{center}
  \includegraphics[width=110mm, angle=270]{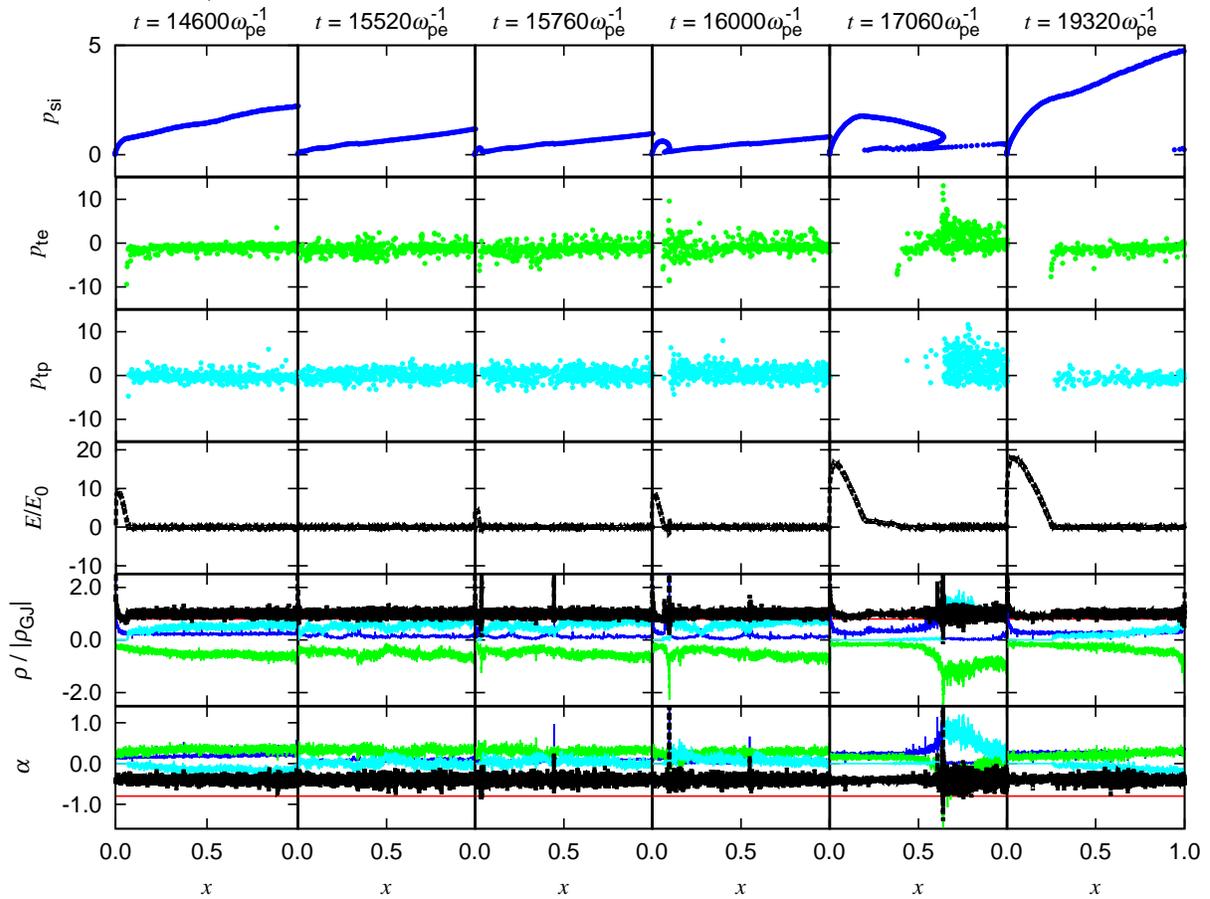}
 \vspace{1cm}
 \caption{The characteristic cycle of the ejection of the particle bunch in Model F2 $(n_{\rm bk, th}/n_{_{\rm GJ}}=0.4)$. 
The length of the calculation domain is $L=10^3\lambda_{\rm pe}$.
The momentum of ions from the NSS is denoted by $p_{\rm si}$ (blue) normalized by $m_{\rm i}c$.
The plotted quantities are the same as in Figure \ref{super05}, 
except for blue lines, which show the contribution of ions from the NSS. 
Note that the time intervals between the snapshots are not equal.} 
 \label{super15}
 \end{center}
\end{figure*}
\begin{figure}
 \begin{center}
  \includegraphics[width=85mm]{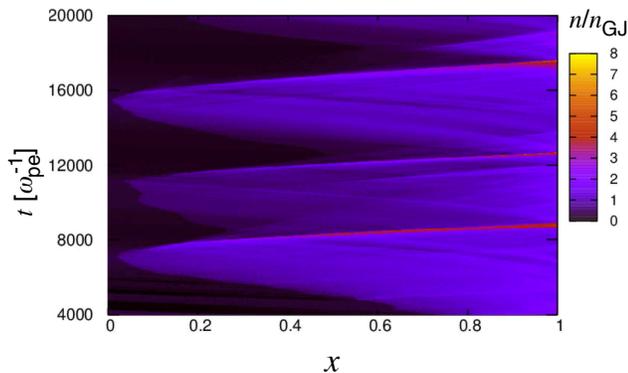}
 \caption{The temporal evolution of the number density of the thermally injected component
   for Model F2 ($n_{\rm bk, th}/n_{_{\rm GJ}}=0.4$). 
The horizontal axis is the distance $x$ and the vertical axis is the time $t$.}
 \label{evolution}
 \end{center}
\end{figure}

Although we confirm that the screening condition of the electric field (equation \ref{screen1}) 
also works in the ion-extracted case,
in some cases we obtain the quasi-periodic solutions,
which are not seen in the electron-extracted cases. 
The typical one cycle of the quasi-periodic behavior for Model F2 
$(n_{\rm bk, th}/n_{\rm GJ}=0.4)$ are shown in Figure \ref{super15}. 
Initially (see the first row) the thermal component from the outer boundary already reaches to the NSS 
and screens the electric field over the entire calculation domain. 
During the electric field is screened out, newly extracted ions are not significantly accelerated ($\ll c$) 
and the current density $\alpha_{\rm ns}$ is suppressed (see the second row).
In order to maintain the condition $\alpha=\alpha_{\rm m}$,
the electric field near the NSS starts to increase and ions are extracted from the NSS (see the third row). 
However, since the ion inertia is larger compared to the electron one,
ions are not immediately accelerated and the total current density $\alpha$ does not reach $\alpha_{\rm m}$ at that time. 
Then, the electric field is continuously grown owing to the difference $\alpha-\alpha_{\rm m}$. 
Although the thermal electrons are accelerated toward the NSS and positrons are reflected by the electric field, 
the electron/positron number density at the NSS is not enough to screen the electric field. 
Subsequently, the electric field significantly develops at the NSS (see the fourth row). 
When the electric field develops enough to accelerate ion up to relativistic velocity, 
accelerated ions catch up to slowly moving ions ejected earlier and a bunch of ions 
(i.e., over-density region) is generated (see the fifth row). 
This ion bunch reflects positrons and traps electrons in order to adjust the charge density 
$\rho\rightarrow\rho_{\rm GJ}$. 
This process increases the number density of the particles in the bunch. 
We show the evolution of the number density distribution for the thermally injected electrons/positrons
in Figure \ref{evolution}. 
Finally, the bunch escapes from the calculation domain (see the sixth row in Figure \ref{super15}) 
and again the thermal component starts to screen the electric field 
in the calculation domain. 
This sequence of the instability occurs quasi-periodically as shown in Figure \ref{evolution}.

\begin{figure}
 \begin{center}
  \includegraphics[width=60mm, angle=270]{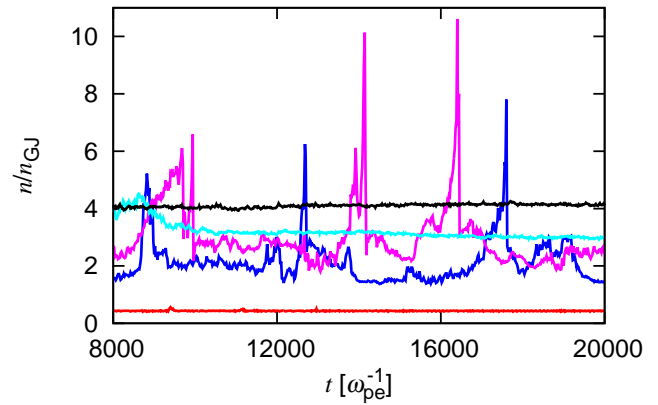}
 \caption{The number density of particles except for the beam component from the OAR 
at the outer boundary in Models F1-5 as a function of the time $t$. 
The different colors show  for $n_{\rm bk, th}/n_{\rm GJ}=$ 0.2 (red; Model F1), 0.4 (blue; Model F2), 
0.6 (magenta; Model F3), 1.1 (cyan; Model F4) and 1.6 (black; Model F5).}
 \label{t-n15}
 \end{center}
\end{figure}

The quasi-periodic solutions are obtained only in limited parameter ranges. 
Figure \ref{t-n15} shows the temporal evolutions of the density of particles 
except for the beam component from the OAR at the outer boundary 
in Models F1-5. 
The quasi-periodic behavior is only seen for Models F2 $(n_{\rm bk, th}/n_{\rm GJ}=0.4$; blue) 
and F3 $(n_{\rm bk, th}/n_{\rm GJ}=0.6$; magenta). 
Since Model F1 
$(n_{\rm bk, th}/n_{\rm GJ}\le0.2$; red) is not screened the electric field over the calculation domain, 
the thermal component develops to the beam particles. 
This solution is almost the same one in Model C2.
If a too large number density of the thermal particles are injected at the outer boundary
($n_{\rm bk,th}/n_{\rm GJ}\gtrsim1.0$, Models F4 and F5), $\alpha_{\rm ns}\rightarrow0$
and ions are not extracted from the NSS.
Then, the solution becomes quasi-steady. 
Therefore, the quasi-periodic solution is obtained in the cases with $\alpha_{\rm ns}\neq0$. 
This means that the cases with the R-sub-GJ and R-anti-GJ conditions possibly have the quasi-periodic solutions 
in some ranges of the density of the thermal component. 
On the other hand,
the R-anti-GJ case with $-1<\beta_{\rm ns}^{\rm req}\le0$ is not expected to have the quasi-periodic solution, 
because the screened state of the electric field is always $\alpha_{\rm ns}=0$
as shown in the electron-extracted case (Figure \ref{anti11_2}).

\begin{figure}
 \begin{center}
  \includegraphics[width=60mm, angle=270]{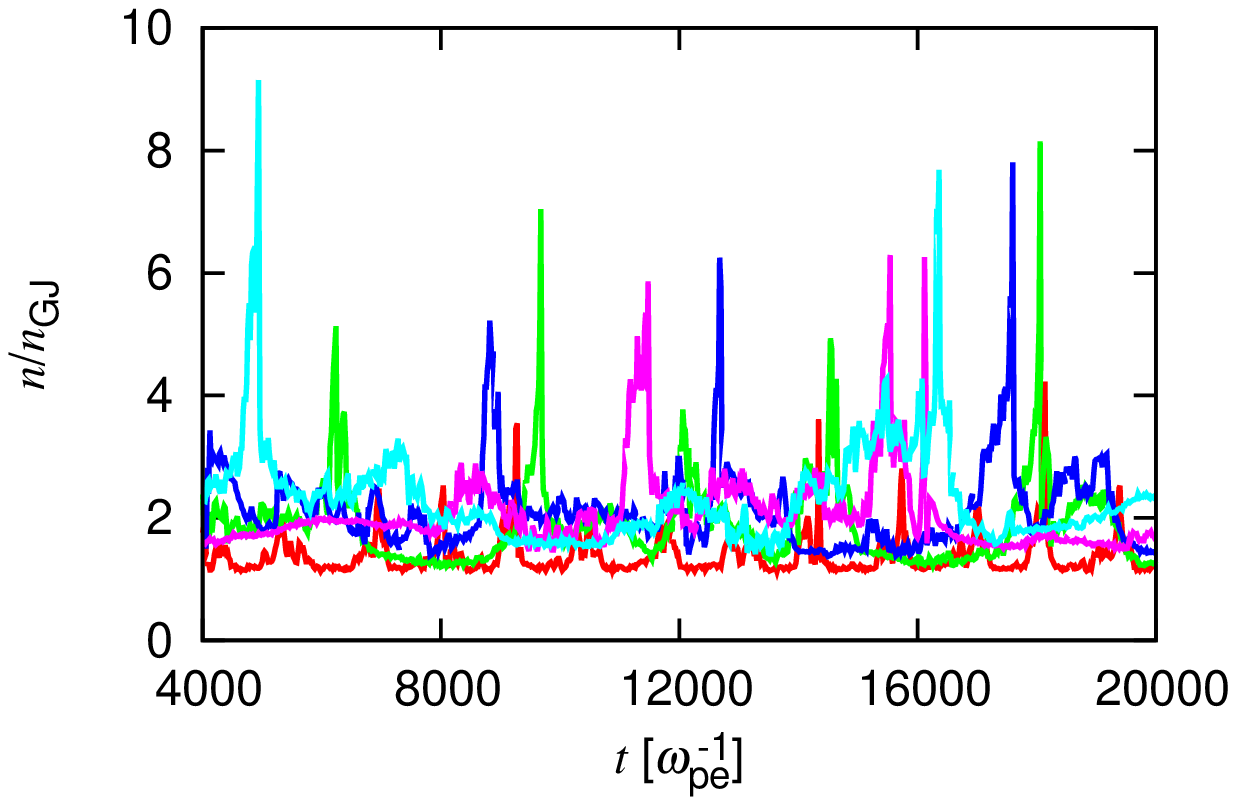}
 \caption{Same as Figure \ref{t-n15} but
   for different computational lengths $L$; 
 $L/\lambda_{\rm pe}$ are 100 [red; Model F2(a)], 500 [green; Model F2(b)], 1000 [blue; Model F2(c)], 2000 [magenta; Model F2(d)] and 3000 [cyan; Model F2(e)].}
 \label{t-n20}
 \end{center}
\end{figure}
\begin{figure}
 \begin{center}
  \includegraphics[width=75mm, angle=270]{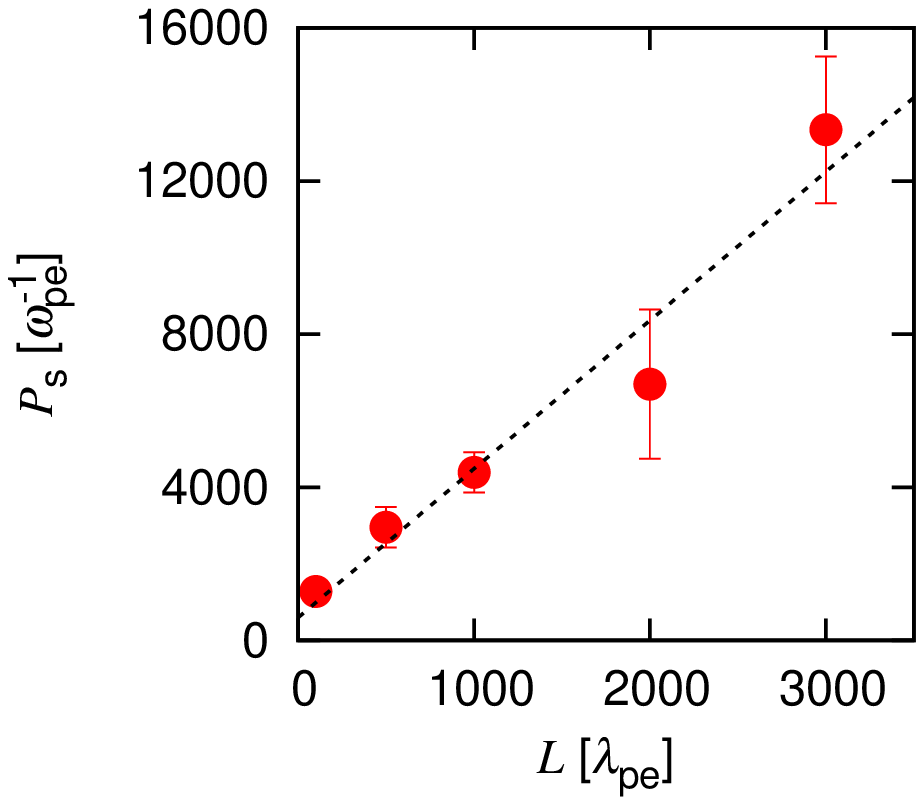}
 \caption{The relation between the average period of the bunch ejection and the scale length of the computational domain 
in Models F2(a-e). The dashed line corresponds to equation (\ref{p_ins}).} 
 \label{P-L}
 \end{center}
\end{figure}

Although the qualitative results in the quasi-steady state (e.g., Models F4 and F5) do not depend on 
the length of the calculation domain $L$,
the period of the sequence and the growth of the bunch may depend on $L$. 
We perform simulations with different domain length $L$. 
In all the cases, we fix the current densities and the injected number density as Model F2.
We labels as Models F2(a) ($L=100\lambda_{\rm pe}$), F2(b) ($L=500\lambda_{\rm pe}$), 
F2(c) ($L=1000\lambda_{\rm pe}$), F2(d) ($L=2000\lambda_{\rm pe}$), and F2(e) ($L=3000\lambda_{\rm pe}$). 
Figure \ref{t-n20} shows the temporal evolution of the number density of particles
except for the beam component from the OAR 
at the outer boundary for various $L/\lambda_{\rm pe}$. 
Except for Model F2(a),
the peak number density is about an order of magnitude larger than the number density 
of the thermal component at the injection. 
The period of the sequence $P_{\rm s}$ clearly depends on $L$. 
In Figure \ref{P-L} we plot the relation between $P_{\rm s}$ and $L$. 
For Models F2(a-e), we obtain the relation between $P_{\rm s}$ and $L$ as 
\begin{eqnarray}\label{p_ins}
P_{\rm s}\sim 3.9L/c+6.0\times10^2\omega_{\rm pe}^{-1}.
\end{eqnarray}
The timescale for the propagation of a bunch of particles is almost a few times of the light crossing time. 
The intercept in equation (\ref{p_ins}) is almost consistent with the acceleration time of ions, 
$t\omega_{\rm pe}\sim(m_{\rm i}/m_{\rm e})\sim 10^3$ for $E/(m_{\rm e}c\omega_{\rm pe}/e)\sim2$. 

\section{Implications for Radio Emission} 
\label{radio}

The coherent curvature radiation from the bunch of particles
has been considered to be a possible mechanism of the observed pulsar radio emission 
\citep[e.g., ][]{RS75}. 
We estimate the luminosity of the radio emission based on our results in the ion-extracted case,
where the bunch of particles is formed quasi-periodically. 
From the simulation results, the size of a bunch is an order of $\sim 10\lambda_{\rm pe} \sim 20$ cm and
the number density of the bunching particles is several times of the GJ number density. 
The period of the quasi-periodic solution is an order of the light crossing time of the calculation domain.

In our results in the previous section, the Lorentz factor of the bunch particles is up to $\gamma\sim 10$. 
However, this value should depend on the number density of extracted ions.
In some parameter sets we performed, 
the bunching particles are heated up to $\gamma\sim10^2$. 
The generating electric field near the NSS accelerates ions up to relativistic velocity in order to adjust the current density. 
Then, the maximum Lorentz factor of the electron/positron, which is determined by the potential energy near the NSS, 
can be an order of the mass ratio $\sim m_{\rm i}/m_{\rm e}\sim10^3$.

The number of particles in a bunch is 
\begin{eqnarray}
N\sim\kappa n_{\rm GJ}\times(10\lambda_{\rm e})^3,
\end{eqnarray}
where $\kappa$ is the average density in the bunch normalized by $n_{\rm GJ}$. 
We assume that multiple bunches compose a thin shell whose surface area is comparable to the polar cap surface. 
In this case, the number of bunches in a shell is $N_{\rm bunch}\sim 10\lambda_{\rm e}\times\pi r_{\rm pc}^2/(10\lambda_{\rm e})^3$. 
We also assume $P_{\rm s}\sim r_{\rm pc}/c$ so that the number of the shell 
we can observe in one rotational period is $N_{\rm shell}\sim\Delta P/P_{\rm s}$, 
where $\Delta P$ is the observed pulse width and we use $\Delta P\sim10^{-2}P^{1/2}$ 
based on the observational results in \citet{MG11}. 
Then, the coherent radio luminosity is estimated as
\begin{eqnarray}\label{radioluminosity}
L_{\rm radio}&\sim& \dot{E}_{\rm cur}N^2N_{\rm bunch}N_{\rm shell} \nonumber \\
&\sim&10^{28}\left(\frac{\kappa}{10}\right)^2\left(\frac{\gamma}{10^2}\right)^4\left(\frac{R_{\rm cur}}{10^7{\rm cm}}\right)^{-2}\ 
{\rm erg\ s}^{-1},
\end{eqnarray}
where $\dot{E}_{\rm cur}$ is the power of curvature radiation for single particle 
and $R_{\rm cur}$ is the radius of the field line curvature. 
Note that the radio luminosity in equation (\ref{radioluminosity}) does not explicitly depend on the spin-down luminosity. 
The observations also show that the radio luminosity is $\sim10^{27}-10^{31}$ erg s$^{-1}$, regardless of the position 
in the $P$-$\dot{P}$ diagram \citep{Sz14}. The characteristic frequency is 
\begin{equation}
\nu_{\rm c}\sim10^9\ {\rm Hz}\biggl(\frac{\gamma}{10^2}\biggr)^3\biggl(\frac{R_{\rm cur}}{10^7{\rm cm}}\biggr)^{-1}.
\end{equation}
The obtained values are roughly consistent with the observations. 

The important point is that the extraction of ions from the NSS occurs for both sign of $\rho_{\rm GJ}$, 
if the back-flowing particles exist.
Most studies including the ion extraction from the NSS focus on the anti-pulsar ($\rho_{\rm GJ}>0$ at the polar cap) 
\citep[e.g., ][]{CR80, J10, J16}. 
In this paper, we consider the screened state with $\alpha_{\rm ns}\neq 0$. 
The current density of the beam component is either $\beta_{\rm bk, bm}^{\rm req}<-1$ or $\beta_{\rm bk, bm}>0$. 
For $\rho_{\rm GJ}<0$ and $\alpha_{\rm m}<-1$, the ions are extracted from the NSS. 
In this case, the quasi-periodic behavior in Section \ref{ion} is expected to occur. 
For $\rho_{\rm GJ}>0$, 
because of the symmetry to the point $(\alpha_{\rm bk},\alpha_{\rm ns})=(-1,0)$ 
in the $\alpha_{\rm bk}-\alpha_{\rm ns}$ diagram (Figure \ref{diagram}), 
ions are extracted from the NSS for the case $\alpha_{\rm m}>-1$ as discussed in Section \ref{setup}. 
Therefore, the quasi-periodic behavior in Section \ref{ion} is possible for both sign of $\rho_{\rm GJ}$.

\section{SUMMARY} 
\label{summary}

In order to activate the OAR, the electric field along the magnetic field just above the NSS should be screened out.
In this paper, we investigate the condition on the electric field screening just above the NSS with back-flowing particles. 
We have focused on the case so-called anti-GJ condition, which would be established along the magnetic fields 
that connect to the OAR. 
Without the back-flows, the electric field cannot be screened out by the particles extracted from the NSS alone.

First, we consider the case that particles accelerated in the OAR with Lorentz factor $\gamma\gg1$ (the beam component) 
are flowing back to the NSS [Figure \ref{image} (a)].
The current and charge densities of the beam component and particles from the NSS contribute to the total ones.
Then, even in the anti-GJ case, 
there is some screened solutions in a certain region for the combination of the current densities
$\alpha_{\rm m}$ and $\alpha_{\rm bk, bm}$ (Figure \ref{diagram}).
We analytically introduce a parameter $\beta_{\rm ns}^{\rm req}$ 
to characterize the required flow speed of the particles extracted from the NSS.
The electric field is not screened in the conditions so-called R-super-GJ and R-anti-GJ, 
which are defined by $\beta_{\rm ns}^{\rm req}$. 

However, we can expect that a quasi-thermal plasma with a mildly relativistic temperature screens out the electric field 
between the OAR and the outer boundary of the polar cap region [Figure \ref{image} (b)].
Some fraction of such thermal components can also flow back to the NSS.
Using numerical simulations, 
we show that the thermal component can adjust both the current and charge densities to the required values. 
We obtain the minimum number density of the thermal component to screen the electric field in equations (\ref{screen1}) 
and (\ref{screen2}). 

We also investigate the ion-extracted case from the NSS.
The difference of masses of electrons and ions causes 
bunches of particles, which are formed quasi-periodically. 
The period is linearly proportional to the length of the calculation domain.
This may be important process for the coherent radio emission.

\acknowledgments
We are grateful to the anonymous referee for helpful comments. 
We would like to thank Kunihito Ioka, Tsunehiko N. Kato, Yutaka Ohira, Shinpei Shibata and Shuta J. Tanaka 
for fruitful discussions. 
This work is supported by KAKENHI 24103006, 16J06773 (S.K.), 24540258 and 15K05069 (K.A., T.T.).

\end{document}